\documentclass[10pt, journal]{IEEEtran} 
\usepackage{times}
\usepackage{graphicx}
\usepackage{cite}
\usepackage{color}
\usepackage{psfrag}
\usepackage{subfigure}
\usepackage{amssymb}
\usepackage{epsfig}
\usepackage{pifont}
\usepackage{amsmath}
\usepackage{array}
\usepackage{multicol}
\usepackage[T1]{fontenc}
\usepackage{comment}
\usepackage{acronym}
\usepackage{multirow}
\usepackage{epsfig}
\usepackage{booktabs}
\usepackage{subcaption}
\usepackage{caption}
\usepackage{makecell}
\usepackage{graphicx}

\begin{document}

\title{Superimposed-Pilot OTFS Under Fractional Doppler: Modular End-to-End Learning 
}
\author{Yushi Lei, Yusha Liu,  \emph{Member, IEEE}, Guanghui Liu, \emph{Senior~Member, IEEE}, Lei Wan, \emph{Senior~Member, IEEE},  and Kun Yang, \emph{Fellow, IEEE}
\thanks{Yushi Lei, Yusha Liu, and Guanghui Liu are with School of Information and Communication Engineering, University of Electronic Science and Technology of China, Chengdu, 611731, China (emails: yushilei@std.uestc.edu.cn; yusha.liu@uestc.edu.cn; guanghuiliu@uestc.edu.cn;).}
\thanks{Lei Wan is with Xiamen University, Xiamen, China (leiwan@xmu.edu.cn).}
\thanks{Kun Yang is with Nanjing University, Nanjing, China (kunyang@nju.edu.cn).}
}
\maketitle

\begin{abstract}
Orthogonal time frequency space (OTFS) modulation has emerged as a promising candidate to overcome the performance degradation of orthogonal frequency division multiplexing (OFDM), which  are commonly encountered in high-mobility wireless communication scenarios. 
However, conventional OTFS transceivers rely on multiple separately designed signal-processing modules, whose isolated optimization often limits global optimal performance. 
To overcome limitations, this paper proposes a modular deep learning (DL) based end-to-end OTFS transceiver framework that consists of trainable and interchangeable neural network (NN) modules, including constellation mapping/demapping, superimposed pilot placement, inverse Zak (IZak)/Zak transforms, and a U-Net-enhanced NN tailored for joint channel estimation and  detection (JCED), while explicitly accounting for the impact of the cyclic prefix. 
This physics-informed modular architecture provides flexibility for integration with conventional OTFS systems and adaptability to different communication configurations. 
Simulations demonstrate that the proposed design significantly outperforms baseline methods in terms of both normalized mean squared error (NMSE) and detection reliability, maintaining robustness under integer and fractional Doppler conditions. 
The results highlight the potential of DL-based end-to-end optimization to enable practical and high-performance OTFS transceivers for next-generation high-mobility networks.
\end{abstract}


\section{Introduction}
Orthogonal time frequency space (OTFS) modulation has emerged as a highly promising waveform for next-generation high-mobility wireless communication systems. Unlike orthogonal frequency division multiplexing (OFDM), which multiplexes symbols directly in the time–frequency (TF) domain, OTFS places information symbols on a delay–Doppler (DD) grid and performs transmission through a two-dimensional transform between the DD and TF domains \cite{9508932,10453468}. By exploiting the sparsity and structured evolution of wireless channels in the DD domain, OTFS can cast a rapidly time-varying doubly selective channel into an approximately time-invariant representation, thereby improving robustness to Doppler-induced distortions and fast channel dynamics. These properties make OTFS a compelling candidate for emerging sixth-generation (6G) scenarios that demand reliable links under high mobility and pronounced environmental variations, including low Earth orbit (LEO) satellite communications \cite{10685072}, massive multiple-input multiple-output (MIMO) systems \cite{10454001}, and integrated sensing and communications (ISAC) \cite{10404096}.

Existing research on OTFS has focused primarily on channel estimation (CE) \cite{10038843, 10370744}, data detection \cite{9973328,9508141} and joint optimization \cite{9785832, 11071963}.
For CE,
Liu \textit{et al.} \cite{10038843} developed a spectrally efficient CE scheme based on a multi-scattered pilot pattern, complemented by low peak-to-average power ratio (PAPR) placement rules and an iterative data-aided estimator.
Liang \textit{et al.}~\cite{10370744} designed two two-dimensional pilot structures  with  matched filters to significantly reduce pilot overhead and improve CE performance in multi-antenna OTFS.
In terms of data detection, Xiang \textit{et al.} \cite{9508141} proposed a low-complexity message passing-aided detection algorithm. Yuan \textit{et al.} \cite{9973328} presented a novel   scheme, allowing parallel processing of messages while maintaining detection accuracy. 
Beyond separate designs, Wang \textit{et al.} \cite{9785832} proposed a variational Bayesian inference (VBI)-based framework that jointly estimates the channel and detects transmitted symbols, enabling better handling of uncertainty in high-mobility environments.
Gao \textit{et al.} \cite{11071963} proposed a lightweight deep learning (DL) framework that simultaneously estimates the channel and recovers transmitted symbols, using data padding and sliced augmentation to achieve improved performance with low computational complexity.
These studies collectively represent important solution strategies for addressing CE and detection problems in OTFS systems. 

However, the OTFS designs mentioned above assume integer Doppler shifts, which is often unrealistic in practical high-mobility scenarios. In real deployments, Doppler shifts are typically fractional due to continuous motion and time-varying relative velocities. Fractional Doppler breaks the alignment between channel paths and the DD grid, causing energy leakage across Doppler bins and inter-Doppler interference. This leakage degrades channel estimation accuracy and complicates data detection, making fractional-Doppler robustness a key challenge in practical OTFS system design.

Motivated by the Doppler leakage and inter-Doppler interference induced by fractional Doppler, several approaches have been proposed to enhance OTFS robustness, ranging from pilot design to interference-mitigation receivers.
For example,
He \textit{et al.} \cite{10637960} proposed a windowed pilot-aided method that suppresses data interference and derived the Cramér-Rao lower bound (CRLB) for non-rectangular windows, significantly improving estimation accuracy and spectral efficiency. 
Wei \textit{et al.} \cite{9738478} addressed the off-grid problem using sparse Bayesian learning (SBL) to accurately estimate channel parameters located at non-integer DD grid points. 
Similarly, Tang \textit{et al.} \cite{10506450} employed Bayesian compressive sensing to mitigate errors caused by fractional delay and fractional Doppler, effectively reducing model mismatch.
By employing DL, 
Mattu \textit{et al.} \cite{10439989} first applied the least squares (LS) method to obtain an initial channel estimate in the time–frequency domain, followed by a deep neural network (DNN) to refine the preliminary results. 
Likewise, Qing \textit{et al.} \cite{10816508} extracted channel features using differential modulation and decision feedback techniques, which were subsequently augmented with a lightweight NN to improve estimation accuracy.
Recently, purely DL-based methods have been proposed to directly learn the mapping from received pilot signals to channel parameters. 
Guo \textit{et al.} \cite{10138432} designed a DNN that predicts channel gains, delays, and fractional Doppler shifts from pilot signals. 
Ying \textit{et al.} \cite{10747184} introduced a multi-stage hybrid network using autoencoders (AE) and recurrent neural networks (RNN) to capture channel path variations and locate significant coefficients. 
In addition, Qi \textit{et al.} \cite{10856399} developed a deep residual attention network (DRAN), incorporating residual, channel attention (CAM), and positional attention modules (PAM) to reduce interference between data and pilot signals, further enhancing estimation performance.

 For data detection in OTFS systems, traditional approaches have been extensively studied. 
Qu \textit{et al.} \cite{9285313} developed a channel equalizer based on least-squares minimum residual (LSMR) and designed an iterative interference cancellation loop to mitigate multiuser interference and improve symbol recovery reliability. 
Similarly, Raviteja \textit{et al.} \cite{8424569} analyzed the interference model in the DD domain and proposed an iterative message-passing (MP) detection algorithm, enhancing detection efficiency through successive refinement.
Hybrid DL approaches have been explored to further improve detection performance. Li \textit{et al.} \cite{10105487} unfolded the classical Expectation Propagation (EP) iterative algorithm into an NN with trainable parameters, combining the structural advantages of traditional algorithms with the flexibility of DL. 
Yue \textit{et al.} \cite{10423015} applied a similar unfolding approach to the Orthogonal Approximate Message Passing (OAMP) algorithm, also incorporating trainable parameters to enhance detection accuracy. 
Furthermore, \cite{10623419} proposed a two-dimensional reservoir computing (2D-RC) NN, which embeds the OTFS physical model via circular padding and filtering, enabling efficient online symbol detection.

Zhang \textit{et al.}~\cite{11263980} proposed a DL-based plug-and-play (PnP) receiver framework that unifies OTFS channel estimation and symbol detection by coupling model-based data-fidelity steps with learnable priors, explicitly targeting scenario-dependent sparsity variations and improving robustness across diverse channel conditions
 
While all the aforementioned methods provide effective solutions for receiver design in OTFS systems, another promising line of research explores end-to-end learning for OTFS transceiver optimization. Unlike only receiver designs, end-to-end training enables joint optimization of the entire transceiver pipeline, potentially capturing DD-domain characteristics more effectively and achieving superior system-level performance.
Singh \textit{et al.} \cite{10552800} employed an AE architecture comprising a DL-based encoder and decoder, trained in an end-to-end manner. However, the underlying model is built upon abstract theoretical formulations without modular design.
Similarly, Tek \textit{et al.} \cite{10226266} trained an AE over an additive white Gaussian noise (AWGN) channel, learning effective mapping and demapping functions. This approach improved bit error rate (BER) performance while reducing detection complexity.
More recently, Cheng \textit{et al.} \cite{11207204} proposed a pilot-free OTFS transceiver using large language models (LLMs), combining artificial intelligence (AI)-driven constellation shaping and LLM-enhanced neural receivers for implicit DD-domain detection, improving spectral efficiency and BER.

Beyond OTFS-specific designs, related modular end-to-end learning approaches in other communication domains provide additional insights.
Zhao \textit{et al.} \cite{9448141} proposed a deep complex convolutional network (DCCN) that replaces the discrete Fourier transform (DFT) and inverse discrete Fourier transform (IDFT) with learned linear transformations, enabling bit recovery from time-domain OFDM signals without explicitly performing DFT/IDFT. Liu \textit{et al.} \cite{liu2022learning} introduced modular models for joint pilot design and precoding.
End-to-end design of polar-coded system is investigated in \cite{hu2024end}, demonstrating BER improvements under perfect channel assumption.

Most existing end-to-end learning approaches for OTFS transceiver design rely on autoencoder-type architectures, in which the transmitter and receiver are jointly optimized as black-box neural networks~\cite{10552800,10226266,11207204}.
Although such designs can achieve performance gains, they suffer from limited interpretability, poor modular reusability, and weak compatibility with conventional OTFS signal processing blocks.

Moreover, most prior end-to-end OTFS designs either assume idealized channel conditions or implicitly absorb channel effects into the neural receiver, without explicitly modeling the impact of fractional Doppler and cyclic prefix (CP) on delay–Doppler domain signal coupling. This limits their applicability in realistic high-mobility scenarios, where fractional Doppler induces severe inter-Doppler interference and degrades the sparsity and structure of the equivalent channel.

To bridge this gap, we propose a modular end-to-end OTFS transceiver framework in which each neural block mirrors a physically meaningful OTFS processing operation, including constellation mapping/demapping, Zak/IZak transforms, and DD-domain processing, while a dedicated receiver module performs joint channel estimation and detection (JCED) with explicit consideration of CP and fractional Doppler effects.

The main contributions of this work are summarized as follows:
\begin{itemize}  
\item{We propose a modular DL-based end-to-end OTFS transceiver framework, in which each NN module is explicitly aligned with a physically meaningful OTFS processing operation. This design preserves the mathematical structure of conventional OTFS processing while enabling global end-to-end optimization across the entire transceiver chain.}

\item{
{More importantly, we design and refine each module in an explainable manner by aligning the neural components with their corresponding conventional signal processing operations. 
Such a modular structure allows each block to be flexibly replaced or integrated with standard OTFS processing units, thereby facilitating practical deployment.} For the conversion between the DD domain and the TF domain, we propose the corresponding DL-IZak/Zak transformation based on mathematical formulas, which not only makes the module interpretable but also provides gains through end-to-end training.}

\item{By explicitly incorporating CP effects and fractional Doppler into the end-to-end learning framework, the proposed architecture enables joint optimization of channel estimation and data detection under realistic doubly dispersive channels. Accordingly, we develop a JCED module based on an enhanced U-Net architecture. Unlike black-box autoencoder designs, the proposed modular framework offers improved interpretability, enhanced flexibility for hybrid deployment with traditional OTFS components, and superior robustness against fractional Doppler-induced interference.
}

\item{Simulation results show that our proposed method achieves superior CE accuracy and improved detection performance under both integer and fractional Doppler conditions. Moreover, the end-to-end joint optimization further enhances the overall detection capability of the system, demonstrating the effectiveness and advantages of the proposed design.}
\end{itemize}

The rest of the paper is organized as follows. Section II describes the OTFS system model from the transmitting end to the receiving end. Section III introduces the fundamental modules in the OTFS transmitter and their corresponding processing components at the receiver. The proposed JCED method is provided in Section IV, while Section V provides the simulation results and performance analysis. Finally, Section VI provides conclusions.

\emph{Notations:}
In this paper, bold uppercase and lowercase letters, such as $\mathbf{X}$ and $\mathbf{x}$, denote matrices and vectors, respectively. 
The notation $\mathbb{C}^{m \times n}$ denotes a complex matrix of dimensions $m \times n$ or a complex vector when $n=1$, while $\mathbb{R}^{m \times n}$ denotes the corresponding real-valued case.
$\mathcal{R}(\cdot)$ and $\mathcal{I}(\cdot)$ indicate the real and imaginary components of a complex vector or matrix.
Matrix inversion, transpose, and Hermitian transpose are represented by $(\cdot)^{-1}$, $(\cdot)^{T}$, and $(\cdot)^{H}$, respectively. 
The absolute value and Euclidean norm are denoted by $\left| \cdot \right|$ and $\| \cdot \|$. 
The expectation operator is given by $\mathbb{E}[\cdot]$, and $\textrm{vec}(\cdot)$ denotes the vectorization of a matrix by stacking its columns (or rows). 
Additionally, the Kronecker product is denoted by $\otimes$. 
$\mathbf{I}_n$ denotes the unit matrix of dimension $n$.

\section{OTFS System Model with SP}
In this section, we present the description of the OTFS transceiver system architecture based on a SP scheme. The system components including the transmitter, channel, and the receiver are introduced sequentially following the natural flow of the communication link. Furthermore, the overall input-output relationship of the transceiver is abstracted and simplified into a matrix formulation, providing a clear mathematical framework for subsequent system analysis and implementation.

\subsection{Transmitter}
\begin{figure}
    \centering
    \includegraphics[width=0.999\linewidth]{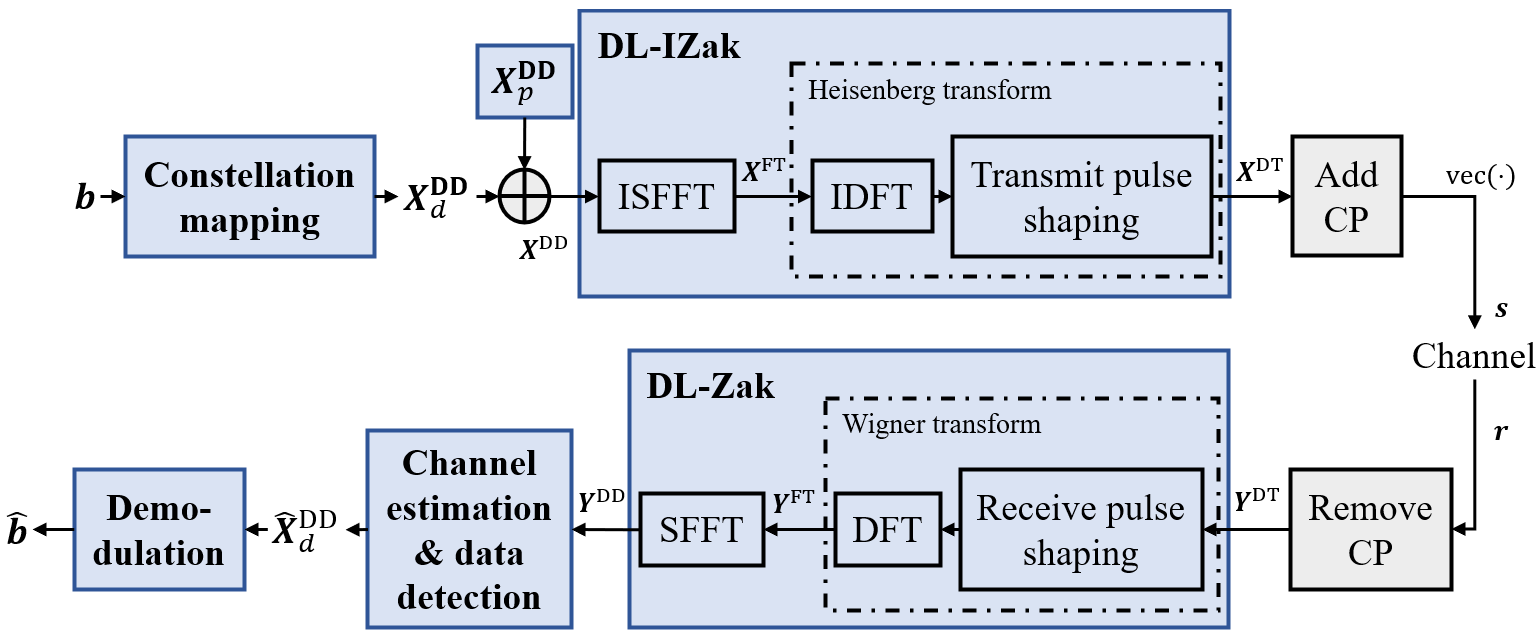}
    \caption{OTFS system model.}
    \label{fig:system model}
\end{figure}

\subsubsection{DD-Domain Mapping}
A basic OTFS system is considered in this paper, as illustrated in Fig.~\ref{fig:system model}, where the transmitted symbols are placed on an $M \times N$ DD-domain grid denoted by $\mathbf{\Gamma}$, which is formally represented as follows:
\begin{align}
\label{eq.grid}
\mathbf{\Gamma} = \Bigg\{ 
\left( \frac{l}{M\Delta f}, \frac{k}{NT} \right) , 
&\, l=0,1,\cdots,M{-}1, \nonumber\\
&\, k=0,1,\cdots,N{-}1
\Bigg\}
\end{align}
where $M$ and $N$ denote the number of delay bins and subcarriers, respectively. The parameters $T$ and $\Delta f$ represent the symbol duration and frequency spacing. The Doppler resolution is given by $\frac{1}{N T}$, while the delay resolution is $\frac{1}{M \Delta f}$. The indices $l$ and $k$ correspond to the discrete delay and Doppler taps, respectively.


Each transmitted data symbol, denoted by $x_d[l,k]$ for $l=0,\ldots,M-1$ and $k=0,\ldots,N-1$, is mapped onto the DD grid defined in (\ref{eq.grid}).

\subsubsection{SPs}
The data mapped onto the DD-domain grid is superimposed with a pilot symbol $x_p$. The resulting composite signal can be represented as
\begin{align}
    {x}[l,k]=\left\{
\begin{array}
{cc}x_p+x_d\left[l_p,k_p\right], & l=l_p,k=k_p, \\
x_d[l,k], & \text{otherwise,}
\end{array}\right.
\end{align}
which can be represented in matrix form as
\begin{align}
\label{eq.pilot}
    \mathbf{X}^{\mathrm{DD}} = \mathbf{X}_d+\mathbf{X}_p,
\end{align}
where $\mathbf{X}^{\mathrm{DD}}$, $\mathbf{X}_d$, $\mathbf{X}_p \in \mathbb{C}^{M \times N}$ denote the DD-domain symbol matrices of the composite signal, data and pilot components, respectively.
The corresponding vectorized forms are given by
\begin{align}
    \mathbf{x}^\text{DD}=\text{vec}(\mathbf{X}^\text{DD}), \quad
    \mathbf{x}_d=\text{vec}(\mathbf{X}_d), \quad
    \mathbf{x}_p=\text{vec}(\mathbf{X}_p),
\end{align}
such that
\begin{align}
    \mathbf{x}^\text{DD}=\mathbf{x}_d + \mathbf{x}_p.
\end{align}
Note that the energy of pilot symbol is denoted by $E_p = |x_p|^2$ and the
average data symbol energy is denoted by $E_d = \mathbb{E} \{|x_d[k,l]|^2\}$.

\subsubsection{TF-Domain Transmission}
The OTFS system employs the inverse symplectic finite Fourier transform (ISFFT) to map the DD-domain signal onto the time-frequency (TF) domain, which is represented as
\begin{align}
\label{eq.DD-TF mapping}
    X[m,n]=\frac{1}{{MN}}\sum_{l=0}^{M-1}\sum_{k=0}^{N-1}x[l,k]e^{-j2\pi\left(\frac{ml}{M}-\frac{nk}{N}\right)}.
\end{align}
In this paper, we adopt the engineering convention for the ISFFT, where the normalization factor $\frac{1}{MN}$ is applied only in the inverse transform. Thus, in matrix form, (\ref{eq.DD-TF mapping}) can be written as
\begin{align}
    \mathbf{X}^{\mathrm{TF}} =\frac{1}{M} \mathbf{F}_M \mathbf{X}^{\mathrm{DD}} \mathbf{F}_N^H,
\end{align}
where $\mathbf{F}_M \in \mathbb{C}^{M \times M}$ and $\mathbf{F}_N \in \mathbb{C}^{N \times N}$ are DFT matrices, with the convention $\mathbf{F}_N^{-1}=\frac{1}{N}\mathbf{F}_N^H$.

Subsequently, the elements of $\mathbf{X}^{\mathrm{TF}}$ are transformed into the time domain through the Heisenberg transform, with the transmit pulse $g_{tx}(t)$ functions serving as the shaping pulse. This operation produces the continuous-time signal $s(t)$, which can be formally expressed as follows
\begin{align}
    s(t) = \sum_{m=0}^{M-1} \sum_{n=0}^{N-1} X[m,n]e^{j 2\pi m \Delta f (t-nT)} g_{tx}(t-nT).
\end{align}
The Heisenberg transform can be regarded as an IDFT operating along the frequency dimension, given by
\begin{align}
    \mathbf{S}=\mathbf{P}_{\text{tx}}\mathbf{F}^H_M \mathbf{X}^\mathrm{TF},
\end{align}
where $\mathbf{S} \in \mathbb{C}^{M \times N}$ is the matrix form of $s(t)$, $\mathbf{P}_{\text{tx}}\in \mathbb{C}^{M \times M}$ is the diagonal transmit pulse-shaping filter matrix assuming a non-ideal rectangular pulse-shape.
Note that the combination of the ISFFT and the Heisenberg transform admits a compact representation via the inverse Zak (IZak) transform \cite{10769778}, which simplifies the operation to an IDFT along the Doppler axis. The resulting time-domain signal is thus expressed as
\begin{align}
\label{eq.IZak}
    \mathbf{S}=\mathbf{P}_{\text{tx}}\mathbf{F}^H_M \mathbf{X}^\mathrm{TF} =\frac{1}{M} \mathbf{P}_{\text{tx}} \mathbf{F}^H_M  \mathbf{F}_M \mathbf{X}^{\mathrm{DD}} \mathbf{F}_N^H =\mathbf{P}_{\text{tx}} \mathbf{X}^{\mathrm{DD}} \mathbf{F}_N^H.
\end{align}

\subsubsection{CP Addition}
To eliminate inter-frame interference, a CP of length 
$L_\text{CP}$ is appended to signal $\mathbf{S}$, where $L_\text{CP}$ is selected to be greater than or equal to the maximum delay spread of the channel. The  final transmitted waveform after adding CP can be represented as \cite{9473532}
\begin{align}
    \mathbf{S}_\text{CP}=\mathbf{A}_\text{CP} \mathbf{S},
\end{align}
where $\mathbf{A}_\text{CP} =
\begin{bmatrix}
\mathbf{0}_{L_\text{CP}\times(M-L_\text{CP})} & \mathbf{I}_{L_\text{CP}} \\
\mathbf{I}_{M}
\end{bmatrix}\in\mathbb{C}^{(M+L_\text{CP})\times M}$ represents the process of CP addition. 
Finally, the transmitter executes parallel-to-serial conversion by vectorizing $\mathbf{S}_\text{CP}$ as $\mathbf{s}_\text{CP} = \mathrm{vec}(\mathbf{S}_\text{CP}) \in \mathbb{C}^{(M+L_\text{CP})N \times 1}$, and subsequently forwards the resulting signal to the transmission front-end for over-the-air propagation.

\subsection{The Doubly-Dispersive Channel Model}
\label{seq.Doubly-Dispersive Channel Model}
After transmission, the signal propagates through the wireless medium and undergoes various distortions, which can be succinctly characterized by a doubly dispersive channel in DD domain. This DD-domain doubly dispersive channel $h(\tau, \nu)$ can be modeled as
\begin{align}
    h(\tau, \nu) = \sum_{i=1}^P h_i \delta (\tau -\tau_i) \delta (\nu -\nu_i),
\end{align}
where $P$ is the number of multi-path. $h_i$ is a complex value representing the channel gains, which remain constant in a time slot based on the assumption of quasi-static propagation channel. $\delta (\cdot) $ represents the Dirac-delta function. $\tau_i$ and $\nu_i$ denote the delay and Doppler shift of the $i$-th path, with $\tau_i \in [0, \tau_{max}]$ and $\nu_i \in [-\nu_{max}, \nu_{max}]$, respectively. 
The channel response in the DD domain can be well located
on the grid as
\begin{align}
    \tau_i = \frac{l_{\tau_i}}{M\Delta f}, \text{and}\ \ \nu_i=\frac{k_{\nu_i}+\kappa_{\nu_i}}{NT},
\end{align}
where $l_{\tau_i}$ and $k_{\nu_i}$ are the integer indices of the delay and the Doppler bins in DD domain. 

In OTFS systems, the delay dimension typically offers sufficiently high resolution, allowing the delays $\tau_i$ to be well approximated as lying exactly on the integer delay grid. In contrast, the Doppler dimension generally has much lower resolution, so the actual Doppler shifts $\nu_i$ rarely coincide with the integer-spaced Doppler grid points. For instance, if the Doppler grid spacing is $\Delta \nu = 50$ Hz but a user experiences a Doppler shift of 37 Hz, this shift cannot be represented by an integer index. To capture this mismatch, the Doppler index is represented using a fractional value $\kappa_{\nu_i} \in [-\frac{1}{2},\frac{1}{2}]$ \cite{8424569}.

Fractional Doppler naturally arises in practical systems because user mobility and relative motion generate continuous Doppler shifts rather than discrete ones. Even small variations in user velocity or angle of arrival can produce Doppler offsets that fall between the predefined grid points. As a result, fractional Doppler is an inherent feature of OTFS channels and poses significant challenges for accurate CE and reliable data detection.

\begin{figure}[t]
  \centering
  \subfigure[magnitude of integrated Doppler channel]{
    \includegraphics[width=0.2\textwidth]{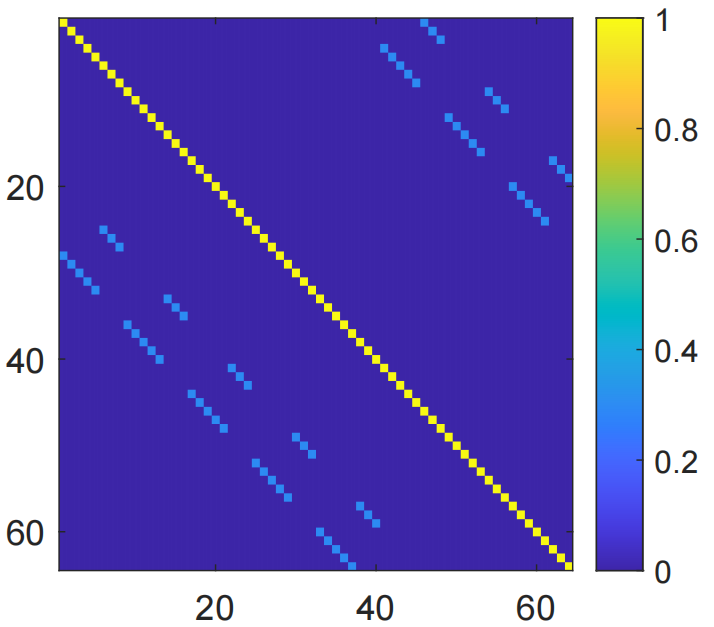}
    \label{fig:channel_int_amp}
  }
  \subfigure[phase of integrated Doppler channel]{
    \includegraphics[width=0.2\textwidth]{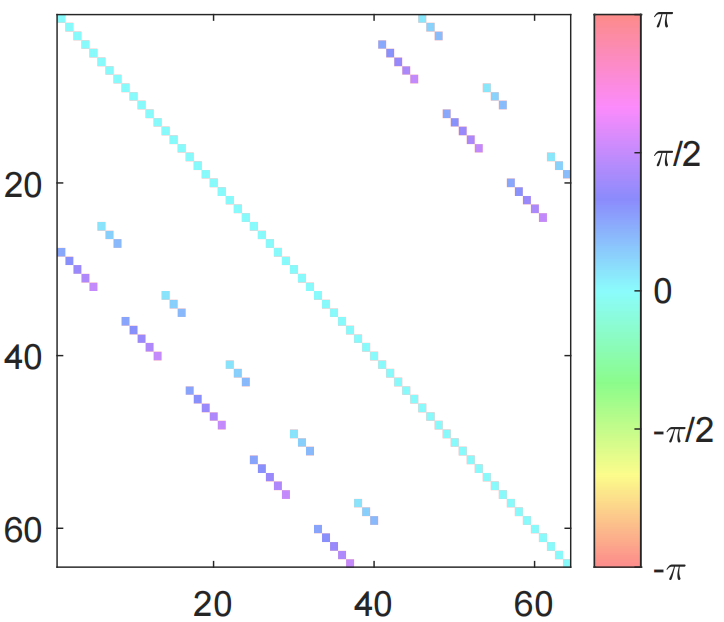}
    \label{fig:channel_int_pha}
  }
  \vspace{0.2cm}
  \subfigure[magnitude of fractional Doppler channel]{
    \includegraphics[width=0.2\textwidth]{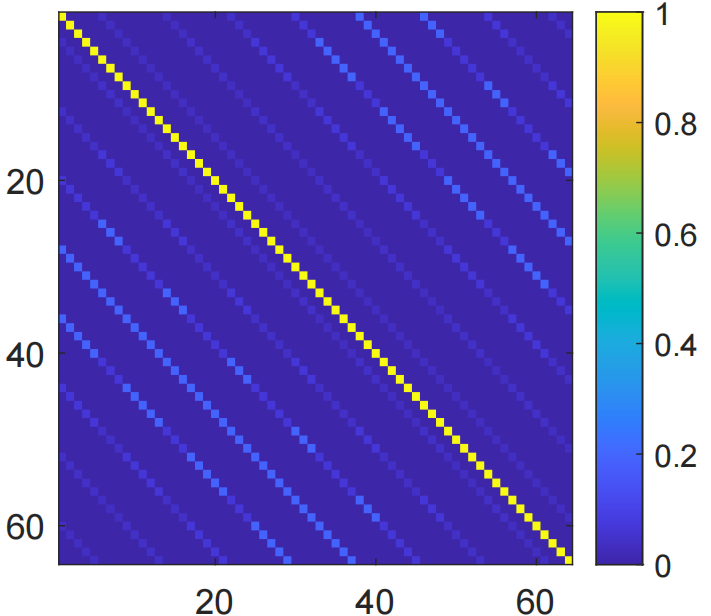}
    \label{fig:channel_frac_amp}
  }
  \subfigure[phase of fractional Doppler channel]{
    \includegraphics[width=0.2\textwidth]{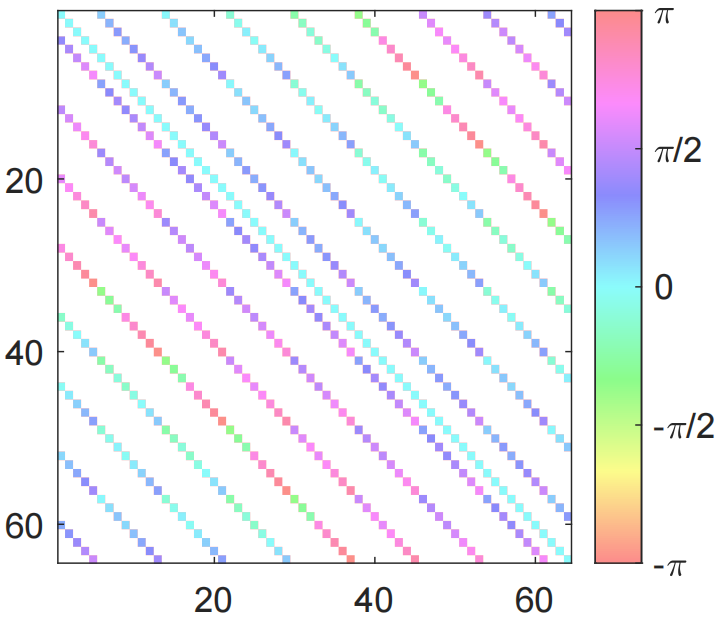}
    \label{fig:channel_frac_pha}
  }
  \caption{Equivalent channel matrices of integer and fractional Doppler.}
  \label{fig:Equivalent channel matrix}
\end{figure}

Fig. \ref{fig:Equivalent channel matrix} illustrates the effect of fractional Doppler on the equivalent channel $\mathbf{H}_\text{eff}$, which is generated according to (\ref{eq.eff channel}) and will be detailed in Section III-C. In this example, $\mathbf{H}_\text{eff}$ comprises two multipath components with delays [0, 3], path gains [1, 0.3], and Doppler shifts set either as [0, 3] for the integer Doppler case or [0, 3.5] for the fractional Doppler case. The corresponding results are shown in Fig. \ref{fig:channel_int_amp} and Fig. \ref{fig:channel_int_pha} for integer Doppler, while in Fig. \ref{fig:channel_frac_amp} and Fig. \ref{fig:channel_frac_pha} for fractional Doppler.

As shown in the figures, fractional Doppler causes energy to leak into neighboring grid points, degrading channel sparsity and resulting in inaccurate CE. 
Ignoring fractional Doppler inevitably yields imprecise CE, which then degrade data detection performance.
This challenge motivates our modular end-to-end transceiver framework, which explicitly accounts for fractional Doppler to enable robust CE and reliable detection under practical high-mobility conditions.

\subsection{Receiver}

At the receiver, the observed signal can be represented as a vector $\mathbf{r}_\text{CP} \in \mathbb{C}^{(M+L_\text{CP})N \times 1}$, corresponding to the sampled version of the transmitted signal vector $\mathbf{s}_\text{CP}$.
In the continuous-time domain, the relationship between the received signal $r_\text{CP}(t)$ and the transmitted waveform $s_\text{CP}(t)$ is given by
\begin{align}
    r_\text{CP}(t)=\int\int h(\tau,\nu)e^{j2\pi\nu(t-\tau)}s_\text{CP}(t-\tau)d\tau d\nu+\omega(t).
\end{align}
Here, the discrete vectors $\mathbf{r}_\text{CP}$ and $\mathbf{s}_\text{CP}$ can be obtained by sampling $r_\text{CP}(t)$ and $s_\text{CP}(t)$ at the receiver, linking the continuous-time formulation to the vector representation used in subsequent processing.

\subsubsection{CP Removal}
The first step in processing the received time-domain signal is to remove the CP of length $L_\text{CP}$. This operation can be denoted in matrix form by 
\begin{align}
    \mathbf{r}=\text{vec}(\mathbf{E}_\text{CP} \mathbf{R}_\text{CP}) ,
\end{align}
where $\mathbf{r} \in \mathbb{C}^{MN \times 1}$ denotes the time domain received signal vector after CP removal, while $\mathbf{R}_\text{CP}\in \mathbb{C}^{(M+L_\text{CP}) \times N}$ is the matrix form of $\mathbf{r}_\text{CP}$. $\mathbf{E}_\text{CP}=[ \mathbf{0}_{M \times L_\text{CP}}, \mathbf{I}_{M}]$ is the CP removal matrix.

\subsubsection{Effective Channel}
The relationship between $\mathbf{r}$ and the corresponding time domain transmitted signal vector $\mathbf{s} = \text{vec}(\mathbf{S})$ with SPs can be succinctly expressed as 
\begin{align}
    \mathbf{r} = \mathbf{H}_{\text{eff}} \mathbf{s} + \mathbf{w},
\end{align}
where $\mathbf{H}_{\text{eff}} \in \mathbb{C}^{MN \times MN}$  denotes the effective channel convolution matrix, and  $\mathbf{w}$  represents the additive noise vector with zero mean and variance $\sigma_n^2$.
Note that $\mathbf{s}$ denotes the vectorized transmit signal before the addition of CP. In other words, the effects of both CP addition matrix $\mathbf{A}_\text{CP}$ and CP removal matrix $\mathbf{E}_\text{CP}$ are incorporated into the equivalent system matrix $\mathbf{H}_\text{eff}$.
The effective channel $\mathbf{H_\text{eff}}$ can be equivalently represented as
\begin{align}
\label{eq.eff channel}
    \mathbf{H}_\text{eff} = \sum_{i=1}^P h_i \mathbf{\Delta}_{k_{\nu_i},l_{\tau_i}} \mathbf{\Pi}_{l_{\tau_i}},
\end{align}
where  $\mathbf{\Delta}_{k_{\nu_i},l_{\tau_i}} \in \mathbb{C}^{MN \times MN}$  accounts for the Doppler shifts, and  $\mathbf{\Pi}_{l_{\tau_i}}\in \mathbb{R}^{MN \times MN}$  corresponds to the delay spread of the channel.
The Doppler shift matrix is a block matrix expressed as
\begin{align}
    \mathbf{\Delta}_{k_{\nu_i},l_{\tau_i}} = \text{diag}[\mathbf{\Delta}_{k_{\nu_i},l_{\tau_i}}^\textbf{(1)},\cdots,\mathbf{\Delta}_{k_{\nu_i},l_{\tau_i}}^\textbf{(N)}],
\end{align}
where 
\begin{align}
\mathbf{\Delta}_{k_{\nu_i},l_{\tau_i}}^{\textbf{(n)}} \triangleq 
\text{diag}\Big[ &\,
w^{(M + L_\text{CP})(n-1) + L_\text{CP} - l_{\tau_i}}, \nonumber\\
&\, w^{(M + L_\text{CP})(n-1) + L_\text{CP} - l_{\tau_i} + 1}, \nonumber\\
&\, \cdots, \nonumber\\
&\, w^{(M + L_\text{CP})n - l_{\tau_i} - 1}
\Big]\in \mathbb{C}^{M \times M},
\end{align}
with $w=e^{\frac{j2\pi (k_{\nu_i}+\kappa_{\nu_i})}{(M+L_\text{CP})N}}$. The delay matrix $\mathbf{\Pi}_{l_{\tau_i}}$ is also a block matrix denoted as
\begin{align}
    \mathbf{\Pi}_{l_{\tau_i}}=\text{diag}[\mathbf{\Pi}_{l_{\tau_i}}^{\textbf{(1)}},\cdots,\mathbf{\Pi}_{l_{\tau_i}}^{\textbf{(N)}}],
\end{align}
where $\mathbf{\Pi}_{l_{\tau_i}}^{\textbf{(n)}} \in \mathbb{C}^{M \times M}$ is the forward cyclic shifted permutation matrix according to delay $l_{\tau_i}$ in the DD domain, i.e. 
\begin{align}
\mathbf{\Pi}_{l_{\tau_i}}^{\textbf{(n)}}=
\left(
\begin{bmatrix}
0 & 0 & \cdots & 0 & 1 \\
1 & 0 & \cdots & 0 & 0 \\
0 & 1 & \cdots & 0 & 0 \\
\vdots & \ddots & \ddots & \ddots & \vdots \\
0 & 0 & \cdots & 1 & 0
\end{bmatrix}_{M\times M}
\right)^{l_{\tau_i}}.
\end{align}

\subsubsection{DD-Domain representation}
The matrix form of the time-domain signal $\mathbf{r}$ is $\mathbf{R} = \text{vec}^{-1}(\mathbf{r}) \in \mathbb{C}^{M \times N}$. This matrix is then mapped to the TF-domain via the Wigner transform, yielding the TF-domain representation
\begin{align}
    \mathbf{Y}^{\text{TF}} = \mathbf{F}_M \mathbf{P}_{\text{rx}}\mathbf{R},
\end{align}
where $\mathbf{P}_{\text{rx}} \in \mathbb{C}^{M \times M}$ denotes the diagonal matched filter matrix of $\mathbf{P}_\text{tx}$.
Next, the TF-domain signal $ \mathbf{Y}^{\text{TF}}$ is mapped to the DD domain via the symplectic finite Fourier transform (SFFT), which is defined as
\begin{align}
    \mathbf{Y}^{\text{DD}} =M\mathbf{F}_M^H \mathbf{Y}^{\text{TF}} \mathbf{F}_N.
\end{align}
Consequently, the DD-domain received vector $\mathbf{y}^\text{DD}\in \mathbb{C}^{MN \times 1}$ has the relationship of
\begin{subequations} \label{eq:vec_y}
\begin{align}
    \mathbf{y}^\text{DD} &= \text{vec}(\mathbf{Y}^{\text{DD}}) \label{eq:vec_y_a} \\
    &= \text{vec}(\mathbf{P}_{\text{rx}}\mathbf{R} \mathbf{F}_N)
    \label{eq:vec_y_b} \\
    &= (\mathbf{F}_N \otimes \mathbf{P}_{\text{rx}})\mathbf{r} \label{eq:vec_y_c} \\
    &= (\mathbf{F}_N \otimes \mathbf{P}_{\text{rx}})\mathbf{H}_{\text{eff}}(\mathbf{F}_N^H \otimes \mathbf{P}_{\text{tx}})\mathbf{x}^\text{DD} 
    + (\mathbf{F}_N \otimes \mathbf{P}_{\text{rx}})\mathbf{w} \label{eq:vec_y_d} \\
    &= \mathbf{G} \mathbf{x}^\text{DD} + \tilde{\mathbf{w}},
    \label{eq:vec_y_e}
\end{align}
\end{subequations}
where the transformation from (\ref{eq:vec_y_b}) to (\ref{eq:vec_y_c}) follows the vectorization property of matrix products, namely that $\text{vec}(\mathbf{AXB})=(\mathbf{B}^T \otimes \mathbf{A})\text{vec}(\mathbf{X})$ \cite{10769778}. The equivalent DD-domain channel matrix $\mathbf{G} \in \mathbb{C}^{MN \times MN}$ is given by
\begin{align}
    \mathbf{G} = (\mathbf{F}_N \otimes \mathbf{P}_{\text{rx}})\mathbf{H}_{\text{eff}}(\mathbf{F}_N^H \otimes \mathbf{P}_{\text{tx}}).
\end{align}
$\tilde{\mathbf{w}} \in \mathbb{C}^{MN \times 1}$ is the effective DD-domain noise vector.

Given that the received DD-domain signal $\mathbf{y}^\text{DD}$ contains contributions from both data symbols $\mathbf{y}_d$ and SPs $\mathbf{y}_p$, further processing is required to isolate these components. 
Specifically, the receiver must first identify the pilot portion embedded within $\mathbf{y}^\text{DD}$, leveraging the known pilot structure for reliable separation. The extracted pilots are then used to estimate the equivalent DD-domain channel $\hat{\mathbf{G}}$, enabling the reconstruction of the channel characteristics experienced by the transmitted symbols. 
With the estimated channel, the pilot-induced interference in $\mathbf{y}^\text{DD}$ is mitigated, producing a refined observation that predominantly contains the data component $\mathbf{y}_d$. Finally,  $\mathbf{y}_d$, together with $\hat{\mathbf{G}}$, is used to detect the transmitted data symbols, as detailed in Section IV.

\section{Proposed OTFS Transceiver based on DL}
\label{sec.Proposed OTFS Transceiver based on DL}

In this section, we propose a DL-based modular OTFS transceiver. 
NN modules are designed to replace key OTFS processing blocks, including constellation mapping, superimposed pilot insertion, Zak/IZak transforms, and JCED. 
This modular structure preserves compatibility with conventional OTFS architectures while enabling flexible integration and system-wide end-to-end optimization. 
Design details for the mapping/demapping, pilot insertion, and Zak/IZak modules are presented in the following subsections, whereas the JCED module will be discussed separately in Section~\ref{sec.JCEDD}.

\subsection{Constellation Mapping and Demapping}

Constellation mapping is the first step in the OTFS modulation process, where the input bit sequence $\mathbf{b}$ is converted into complex-valued modulation symbols for placement on the DD-domain grid.
Traditional fixed constellations, such as binary phase shift keying (BPSK) and quadrature phase shift keying (QPSK), are designed based on idealized channel assumptions and therefore lack the ability to adjust their symbol positions under varying operating conditions. Consequently, when the system experiences different SNR levels or propagation conditions, these fixed constellations may no longer be optimal, leading to sub-optimal BER performance.
To overcome this limitation, we adopt a NN–based trainable constellation mapping strategy, which learns the constellation points in an end-to-end manner \cite{luo2025bedrock}.
By allowing the symbol locations in the complex plane to adjust dynamically during training, the model can tailor the constellation to the target channel characteristics and effectively minimize BER. 

The constellation mapping operation is parameterized by a trainable complex-valued vector set $\mathbf{M}=[{M}_1, {M}_2,\cdots,{M}_R]^T \in \mathbb{C}^{R \times 1}$, where $R = 2^r$ is the modulation order, $r$ denotes the number of bits per symbol, e.g., $r=1$ for BPSK and $r=2$ for QPSK. 
Each entry ${M}_g$ represents the learned constellation point corresponding to the $g$-th symbol.
For each input symbol, its associated bit sequence is first converted into a one-hot vector $\mathbf{e}_g \in \{0,1\}^{1 \times R}$, which selects the corresponding constellation point. The $g$-th mapped complex-valued symbol $x_g$ is then obtained as
\begin{align}
    x_g = \mathbf{e}_g \mathbf{M},
\end{align}
which effectively extracts the $g$-th element ${M}_g$ from the constellation vector $\mathbf{M}$.

To enable stable learning of the constellation mapping process, a power-normalization step is applied during each training iteration. This ensures that the average energy of the learned constellation remains within the desired power constraint. Specifically, each constellation point is rescaled such that the average magnitude of all points equals one. The normalized constellation point is computed as
\begin{align}
    \tilde{{M}}_g = \frac{{M}_g}{\sqrt{\frac{1}{R}\sum_{g=1}^R(\mathcal{R}({M}_g)^{2}+\mathcal{I}({M}_g)^{2})}},
\end{align}
where, $\tilde{\mathbf{M}}_g$ denotes the $g$-th constellation point after normalization, $\mathcal{R}(\cdot)$ and $\mathcal{I}(\cdot)$ denote the real and imaginary parts of a complex number, respectively.

At the receiver, corresponding to the constellation mapping performed at the transmitter, the goal is to recover the original bit sequence from the received DD-domain signal.
To accomplish this, we employ a three-layer fully connected DNN to perform the demapping task. Specifically, the first two layers of the DNN use ReLU activation functions to introduce nonlinearity, while the output layer employs a linear activation to produce the final demapped values.

The input to this DNN is prepared by vectorizing the received complex-valued DD-domain signal $\hat{\mathbf{X}}^\text{DD}_d \in \mathbb{C}^{M \times N}$  into a column vector $\hat{\mathbf{x}}^\text{DD}_d \in \mathbb{C}^{MN \times 1}$, then separating and concatenating its real and imaginary parts to form a real-valued matrix $[\mathcal{R}(\mathbf{x}),\mathcal{I}(\mathbf{x})]\in \mathbb{R}^{MN \times 2}$. 
Each DNN layer is followed by batch normalization, and the network outputs a probability estimate for each bit $\hat{\mathbf{b}}$, completing the demapping process.

\subsection{Superimposed Pilot Insertion}

After constellation mapping, the modulated data symbols are arranged into a complex-valued matrix $\mathbf{X}^\text{DD}_d \in \mathbb{C}^{M \times N}$ , which represents the DD-domain data grid for further processing. A complex pilot symbol $x_p$ is then superimposed onto a designated grid location $(l_p, k_p)$ of $\mathbf{X}^\text{DD}_d$, as defined in (\ref{eq.pilot}).

To improve efficiency, $x_p$ is implemented as a trainable complex neuron. During end-to-end training, this learnable pilot automatically adjusts its phase to improve estimation accuracy.
Unlike conventional designs where the pilot has a fixed value and requires higher transmission power, the proposed adaptive pilot achieves reliable CE at relatively lower power. This flexibility may also help improve PAPR performance, enabling a more power-efficient and hardware-friendly pilot design within the OTFS framework.

\subsection{DL-IZak and DL-Zak}
\begin{figure}[t]
  \centering
  \subfigure[ ]{
    \includegraphics[width=0.23\textwidth]{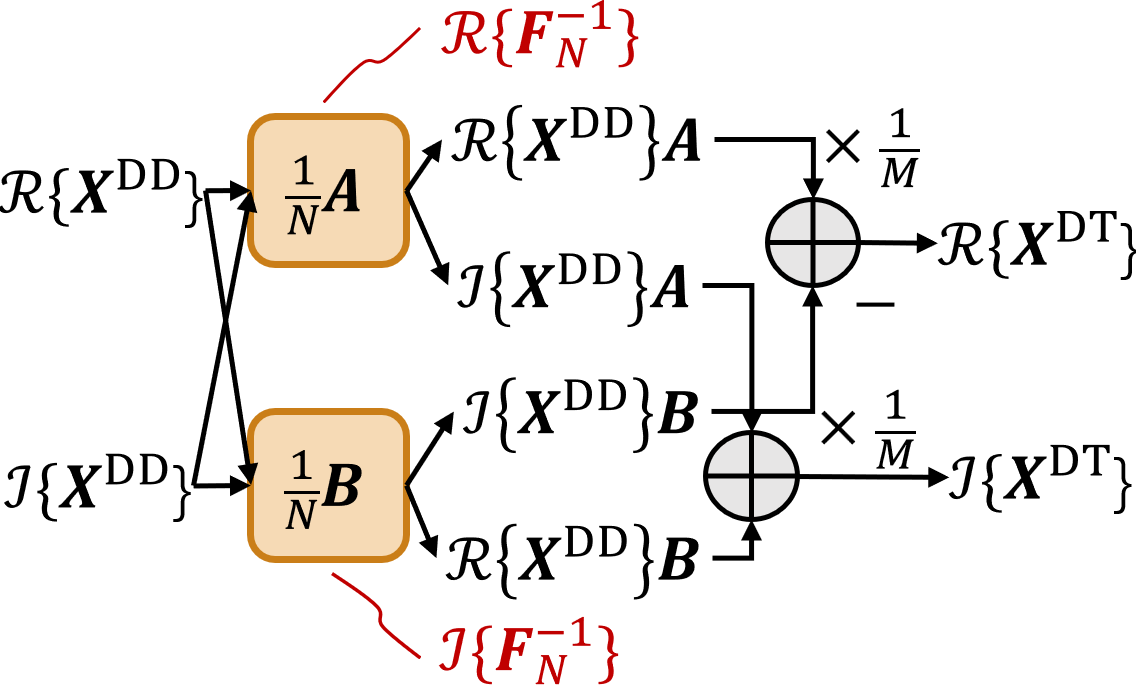}
    \label{fig:IZak}
  }
  \hspace{-4mm}  
  \subfigure[ ]{
    \includegraphics[width=0.23\textwidth]{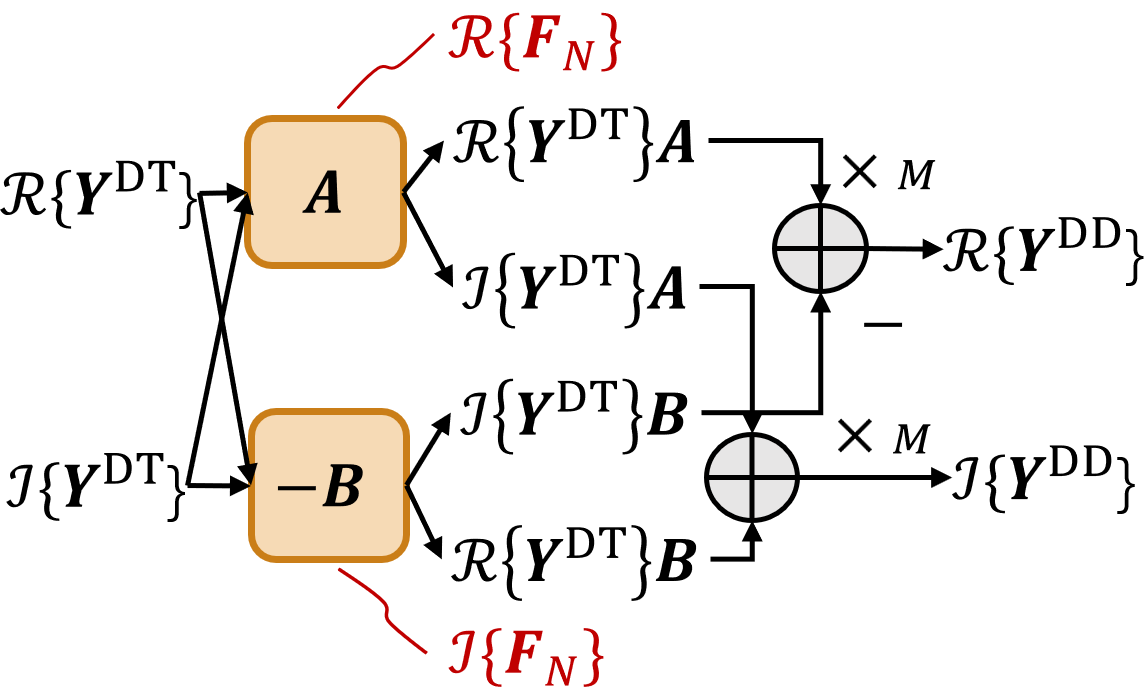}
    \label{fig:Zak}
  }
  \caption{The model design of  (a) DL-IZak and (b) DL-Zak.}
  \label{fig:main}
\end{figure}

The transformation from DD-domain to time-domain is achieved by ISFFT and Heisenberg transform successively.
This entire process can be simplified as the IZak transform along the Doppler axis, as expressed in (\ref{eq.IZak}).
To realize this transformation in a learnable manner, we design a DNN named DL-IZak. Observing that the IZak transform essentially corresponds to a matrix multiplication between $\mathbf{X}^\text{DD}$ and IDFT matrix $\mathbf{F}_N^{-1}$, we model this operation using two trainable real-valued matrices $\mathbf{A}$ and $\mathbf{B}$ and multiply them with $\frac{1}{N}$ to represent the real and imaginary parts of the IDFT matrix, $\mathcal{R}(\mathbf{F}_N^{-1})$ and $\mathcal{I}(\mathbf{F}_N^{-1})$, respectively, as shown in Fig.~$\ref{fig:IZak}$. 
The output of DL-IZak corresponds to the delay-time (DT)-domain signal $\mathbf{X}^\text{DT}$.  The mathematical formulation of DL-IZak can be expressed as 
\begin{align}
    \mathcal{R}(\mathbf{X}^\text{DT}) & =\frac{1}{M}\left(\mathcal{R}(\mathbf{X}^\text{DD})\mathcal{R}(\mathbf{F}_N^{-1})-\mathcal{I}(\mathbf{X}^\text{DD})\mathcal{I}(\mathbf{F}_N^{-1})\right) \nonumber \\
    & = \frac{1}{M}\left(\mathcal{R}(\mathbf{X}^\text{DD})(\frac{1}{N}\mathbf{A})-\mathcal{I}(\mathbf{X}^\text{DD})(\frac{1}{N}\mathbf{B})\right),
\end{align}
\begin{align}
    \mathcal{I}(\mathbf{X}^\text{DT}) & =\frac{1}{M}\left(\mathcal{R}(\mathbf{X}^\text{DD})\mathcal{I}(\mathbf{F}_N^{-1})-\mathcal{I}(\mathbf{X}^\text{DD})\mathcal{R}(\mathbf{F}_N^{-1})\right) \nonumber \\
    & =\frac{1}{M}\left(\mathcal{R}(\mathbf{X}^\text{DD})(\frac{1}{N}\mathbf{B})-\mathcal{I}(\mathbf{X}^\text{DD})(\frac{1}{N}\mathbf{A})\right).
\end{align}

At the receiver, the DT-domain signal $\mathbf{Y}^\text{DT}$ is processed by the Wigner transform followed by the SFFT. To facilitate efficient processing, we introduce a DNN termed DL-Zak. Since the Zak transform essentially corresponds to a matrix multiplication between $\mathbf{Y}^\text{DT}$ and DFT matrix $\mathbf{F}_N$, and because the Zak and IZak transforms are related through conjugate-transpose operations, i.e., $\mathbf{F}_N^{-1}=\frac{1}{N}\mathbf{F}_N^H$, we have
\begin{align}
    \mathbf{F}_N^{-1} &= \frac{1}{N} (\mathbf{A} + j \mathbf{B}), \\
    \mathbf{F}_N &= N (\mathbf{F}_N^{-1})^H = \mathbf{A} - j \mathbf{B},
\end{align}
where $\mathbf{A}$ and $-\mathbf{B}$ can be corresponded to the real and imaginary parts of the DFT matrix, i.e., $\mathcal{R}(\mathbf{F}_N)$ and $\mathcal{I}(\mathbf{F}_N)$, respectively. 
Based on this decomposition, the DL-Zak parameters can be directly derived from $\mathbf{A}$ and $\mathbf{B}$ without additional optimization, reducing training redundancy.
The mathematical formulation of DL-Zak in terms of $\mathbf{A}$ and $\mathbf{B}$ can be expressed as
\begin{align}
    \mathcal{R}(\mathbf{X}^\text{DD}) & =M\left(\mathcal{R}(\mathbf{X}^\text{DT})\mathcal{R}(\mathbf{F}_N)-\mathcal{I}(\mathbf{X}^\text{DT})\mathcal{I}(\mathbf{F}_N)\right) \nonumber \\
    & =M\left(\mathcal{R}(\mathbf{X}^\text{DT})\mathbf{A}+\mathcal{I}(\mathbf{X}^\text{DT})\mathbf{B} \right), 
\end{align}
\begin{align}
    \mathcal{I}(\mathbf{X}^\text{DD})& =M\left(\mathcal{R}(\mathbf{X}^\text{DT})\mathcal{I}(\mathbf{F}_N)-\mathcal{I}(\mathbf{X}^\text{DT})\mathcal{R}(\mathbf{F}_N)\right) \nonumber \\
    &=M\left(-\mathcal{R}(\mathbf{X}^\text{DT})\mathbf{B}-\mathcal{I}(\mathbf{X}^\text{DT})\mathbf{A}\right) .
\end{align} 
The corresponding NN model of DL-Zak is illustrated in Fig.~\ref{fig:Zak}.

\subsection{Training Process}
Since the system takes binary bit sequences $\mathbf{b}\in \{0,1\}$ as input and outputs the corresponding predicted probabilities $\hat{\mathbf{b}}\in (0,1)$ for each bit, the final task can be naturally formulated as a binary classification problem. This formulation allows the NNs to learn to distinguish between bit ‘0’ and bit ‘1’ based on the received signal features.

To train the NN effectively and guide it toward accurate bit prediction, we employ the binary cross-entropy (BCE) loss as the objective function. The BCE loss quantitatively measures the discrepancy between the predicted probabilities and the ground truth bits by penalizing incorrect predictions more heavily. Formally, for a total of $Q$ bits, the BCE loss is defined as
\begin{align}
    \mathcal{L}_{\text{BCE}} = - \frac{1}{Q} \sum_{q=1}^{Q} \left[ b_q \log \hat{b}_q + (1 - b_q) \log (1 - \hat{b}_q) \right],
\end{align}
where $b_q$ denotes the $q$-th ground truth bit, $\hat{b}_q$ is the predicted probability. Minimizing this loss encourages the model to produce probability outputs close to the true bit values, thereby improving demapping accuracy.

For the initialization of the NNs, modules with clear theoretical foundations are initialized to their conventional values. For example, the constellation mapping is initialized to the corresponding BPSK/QPSK constellation points, and the IZak/Zak transform networks are initialized with the values of the corresponding IDFT/DFT matrices. This approach allows our optimization to start from a well-performing baseline, enabling more efficient and effective improvements. All remaining modules without explicit analytical counterparts are initialized randomly following standard NN initialization schemes.

Note that the NN modules introduced above can be trained effectively under a simple AWGN channel. However, real wireless channels exhibit time and frequency selectivity, resulting in doubly dispersive characteristics. In the next section, we extend the system design to this practical setting by introducing a dedicated DL module for JCED. This extension allows the transceiver to handle realistic channel impairments and significantly improves robustness in practical communication environments.

\section{Proposed JCED method}
\label{sec.JCEDD}
\begin{figure*}
    \centering
    \includegraphics[width=0.99\linewidth]{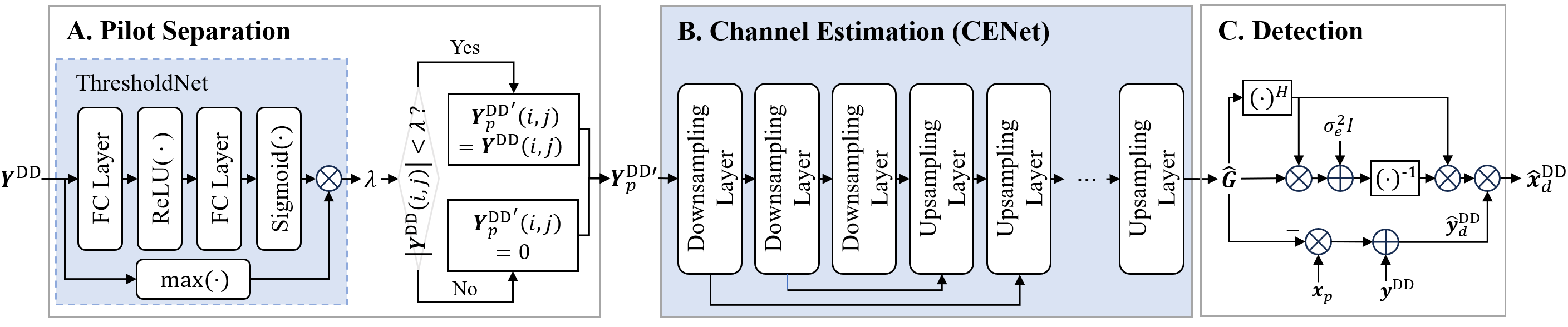}
    \caption{The network structure of the proposed JCED.}
    \label{fig:JCEDD}
\end{figure*}

In this section, we present a detailed design of the receiver modules.    
As shown in Fig.~\ref{fig:JCEDD}, the input to this model    is the received DD-domain signal $\mathbf{Y}^\text{DD}$, and the objective is to estimate the channel and compensate for its effects to recover the transmitted data symbols $\mathbf{X}^\text{DD}_d$.
Since the received signal contains both data and pilot symbols superimposed, the first step is to separate the pilots from the data.
The extracted pilot information is then used for CE, which serves as the foundation for subsequent data detection.
To clearly present the design pipeline, this section is divided into three parts: pilot separation, CE, and detection, which will be discussed in detail in the following subsections.

\subsection{Pilot Separation Network}
To separate pilot symbols from the DD-domain signal $\mathbf{Y}^\text{DD}$, we leverage the fact that pilot symbols are transmitted with a higher power than data symbols. This characteristic allows us to locate pilot positions by identifying high-magnitude entries in $\mathbf{Y}^\text{DD}$.

To automate this process, we propose a lightweight NN module called ThresholdNet, as shown in Fig. \ref{fig:JCEDD}. It takes the magnitude of the received signal $|\mathbf{Y}^\text{DD}|\in \mathbb{R}^{M \times N}$, flattens it into a vector, and feeds it into a two-layer fully connected network: a ReLU-activated hidden layer followed by a Sigmoid-activated output layer that produces a normalized scalar in $(0,1)$. This scalar is then multiplied by the maximum value of $|\mathbf{Y}^\text{DD}|$ to obtain the adaptive threshold $\lambda$, i.e.,
\begin{align}
    \lambda = \text{Sigmoid}\left(\text{FC}_2\left(\text{ReLU}(\text{FC}_1(\text{vec}(|\mathbf{Y}^\text{DD}|)))\right)\right) \times \max (|\mathbf{Y}^\text{DD}|).
\end{align}

We then apply this threshold to the received signal: all entries in $\mathbf{Y}^\text{DD}$ with magnitude below $\lambda$ are set to zero, while entries above the threshold are retained. This results in a sparse $\mathbf{Y}^{\text{DD}'}_p$, which serves as a coarse estimate of the received pilot signal and will be used for subsequent CE.

\subsection{Channel Estimation Network}
In the SP OTFS system, the extracted pilot signals $\mathbf{Y}^{\text{DD}'}_p$  are typically contaminated by overlaid data symbols. 
Note that this interference is non-Gaussian and signal-dependent, while traditional linear estimators such as LS or MMSE struggle to suppress it effectively. As a result, the recovered pilot signal may deviate significantly from its true form, leading to an inaccurately estimated channel and reduced reliability in the subsequent detection stage. Therefore, a more powerful nonlinear model is required to recognize and suppress structured interference while simultaneously extracting meaningful channel features.

\begin{figure*}
    \centering
    \includegraphics[width=0.9\linewidth]{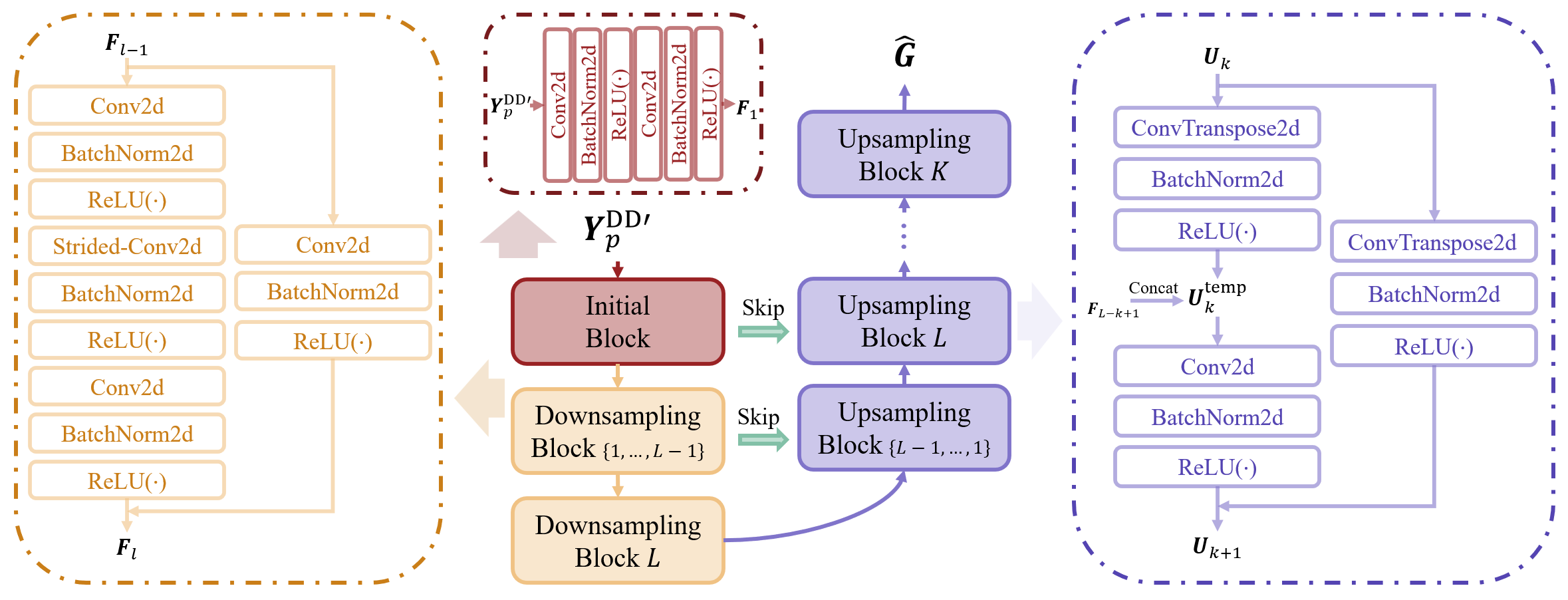}
    \caption{The detailed structure of the CENet.}
    \label{fig:CE}
\end{figure*}

To address this issue, we design a NN-based CE network named CENet, as shown in Fig.~\ref{fig:CE}. The input to CENet is the coarse pilot matrix $\mathbf{Y}_p^{\text{DD}'}$ with the dimension of $[1,M,N]$, while the output is the estimated equivalent DD-domain channel matrix $\mathbf{G}$ with the dimension $[1, M\times N, M \times N]$. 

The core motivation behind CENet is to exploit NNs' ability to learn nonlinear and multi-scale representations, which are essential for handling the interference structure in SP OTFS. In particular, the interference distribution varies across different DD-domain regions, and its correlation structure cannot be accurately described using a shallow model. This motivates the adoption of a multi-resolution learning framework capable of capturing both local DD-domain details and global coupling patterns.

CENet is built upon a modified U-Net architecture and consists of an encoder–decoder-skip structure with residual convolutional blocks \cite{ronneberger2015unetconvolutionalnetworksbiomedical}. The use of U-Net is not arbitrary; rather, it is driven by the specific characteristics of OTFS channels. 
In particular, the encoder progressively downsamples the input to extract increasingly abstract and globally representative features, eliminating the influence of interference.
The encoding process begins with an initialization block
\begin{equation}
    \mathbf{D}_{1} =\mathrm{Int}(\mathbf{Y}_{p}^{\text{DD}'}),
\end{equation}
where $\mathbf{D}_{\ell}$ represents the feature map at the first initialization stage or $\ell$-th downsampling stage with $\ell =2, \cdots, L$. The depth $L$ depends on the dimension of the input matrix $\mathbf{Y}_{p}^{\text{DD}'}$. 
$\mathrm{Int}(\cdot)$ denotes the initial processing illustrated in Fig. \ref{fig:CE}. Then, the $l$-th downsampling block performs
\begin{equation}
    \mathbf{D}_{\ell+1} = \mathcal{E}_{\ell}(\mathbf{D}_{\ell})
    =\mathrm{ResConv}_{\ell} (\mathrm{Down}_{\ell}(\mathbf{D}_{\ell})),
\end{equation}
where $\mathrm{ResConv}_{\ell}(\cdot)$ is a residual convolution block, and $\mathrm{Down}_{\ell}(\cdot)$ is a convolution which reduces the output dimensions to half of the input. 
Through this progressive downsampling, the input resolution is compressed from $[1,M,N]$ to compact forms such as $[64, 2, 2]$. This compression suppresses high-frequency fluctuations introduced by the interference while also forcing the network to learn more abstract patterns of the OTFS channel.

The decoder then progressively upsamples the latent features to reconstruct the full-size DD-domain channel matrix $\hat{\mathbf{G}}$. Formally, the $\kappa$-th upsampling block is defined as
\begin{equation}
    \mathbf{U}_{\kappa+1} = \mathcal{D}_{\kappa}(\mathbf{U}_{\kappa})
    = \mathrm{ResConv}_{\kappa}\!\big(\mathrm{Up}_{\kappa}(\mathbf{U}_{\kappa})\big),
\end{equation}
where $\mathbf{U}_{\kappa}$ is the feature map at the $\kappa$-th upsampling stage with $\kappa =1, \cdots, K$. The value of $K$ depends on the dimension of the output matrix $\hat{\mathbf{G}}$. Here, $\mathrm{Up}_{\kappa}(\cdot)$ represents a transposed convolution block which doubles the output dimensions to twice that of the input, and $\mathrm{ResConv}_{\kappa}(\cdot)$ denotes a residual convolution block. Through this hierarchical reconstruction, the global and abstract features produced by the encoder are transformed back into structures that match the requirements of the DD-domain equivalent channel matrix $\hat{\mathbf{G}}\in \mathbb{C}^{MN \times MN}$.
As $\hat{\mathbf{G}}$ is fed directly to the OTFS equalizer, maintaining its structural consistency with the true channel is essential for correct equalization.

A key component of CENet is the use of skip connections that link encoder and decoder features at corresponding depths. The skip mechanism is defined as
\begin{equation}
\label{eq:skip}
    \mathbf{Z}_{\kappa}
    = \mathrm{Concat}\big(\mathrm{Skip}_{L-\kappa+1}(\mathbf{F}_{L-\kappa+1}),\, \mathbf{U}_{\kappa}^{\text{temp}}\big),
\end{equation}
where $\mathrm{Skip}_{L-\kappa+1}(\cdot)$ adjusts the encoder feature $\mathbf{F}_{L-\kappa+1}$ to match the dimension of $\mathbf{U}_{\kappa}^{\text{temp}}$. In typical configurations, the number of upsampling blocks is larger than that of downsampling blocks, i.e., $K \geq L$. For those additional upsampling layers without corresponding encoder features, no skip connection is used and we simply set $\mathbf{Z}_{\kappa}=\mathbf{U}_{\kappa}^{\text{temp}}$.  
These skip connections effectively restore high-resolution information that may be lost during downsampling thus enhance robustness against interference.

With this encoder–decoder–skip design, CENet captures multi-scale and nonlinear representations that not only suppress data-induced interference but also exploit the inherent structure of OTFS channels to reconstruct a high-fidelity DD-domain equivalent channel matrix $\hat{\mathbf{G}}$, thereby providing a reliable foundation for subsequent data detection.

\subsection{Data Detection Network}

After obtaining the estimated channel matrix $\hat{\mathbf{G}}$, the first crucial step of detection is to remove the influence of the SP signal from the received DD-domain signal $\mathbf{Y}^\text{DD}$, thereby isolating the data component  expressed as
\begin{align}
    \mathbf{y}^\text{DD}_d = \mathbf{y}^\text{DD}-\hat{\mathbf{G}}\mathbf{x}_p,
\end{align}
where $\mathbf{y}^\text{DD}_d$ denotes the vector of data symbols with pilot interference removed, $\mathbf{y}^\text{DD}=\text{vec}(\mathbf{Y}^\text{DD})$, $\mathbf{x}_p = \text{vec}(\mathbf{X}_p)$ is the known pilot vector. This subtraction effectively eliminates the pilot contribution, providing a cleaner data signal for subsequent data detection.

Subsequently, to estimate the transmitted data symbols $\hat{\mathbf{X}}^\text{DD}_d$ from the pilot-free signal $\mathbf{y}^\text{DD}_d$, we employ a linear minimum mean square error (LMMSE) detector. The LMMSE detector achieves a favorable trade-off between computational complexity and detection performance by leveraging both CE and noise statistics. Its detection procedure is given by
\begin{align}
    \hat{\mathbf{x}}^\text{DD}_d = (\hat{\mathbf{G}}^H\hat{\mathbf{G}}+\sigma_e^2\mathbf{I})^{-1}\hat{\mathbf{G}}^H \hat{\mathbf{y}}^\text{DD}_d,
\end{align}
where $\sigma_e^2$ denotes the noise variance. 
Note that the performance of the LMMSE detector critically depends on the accuracy of the channel estimate $\hat{\mathbf{G}}$, any error in $\hat{\mathbf{G}}$ propagates into the detection stage, potentially leading to biased or suboptimal symbol recovery. This strong coupling naturally motivates the design of a unified learning framework, in which CE and data detection are jointly optimized to improve end-to-end performance.

In the proposed design, the LMMSE detector is implemented as a deterministic and fully differentiable layer whose computation depends explicitly on the NN-generated channel estimate $\hat{\mathbf{G}}$. All operations including Hermitian multiplication, matrix inversion, and linear transformation are differentiable, enabling gradients to pass through the detector and flow into CENet. This allows the CENet to not only produce accurate CE but also adapt its output to better support the downstream detection process.

With this architecture, the JCED  module can be seamlessly embedded into the overall OTFS transceiver to enable full end-to-end learning. 
The estimated symbol vector 
$\hat{\mathbf{x}}^\text{DD}_d \in \mathbb{C}^{MN \times 1}$ is reshaped into 
$\hat{\mathbf{X}}^\text{DD}_d \in \mathbb{C}^{M \times N}$, corresponding to the DD-domain data grid, which is then further processed for constellation demapping. 
During training, which is detailed in Section~III-D, the gradients from the detection loss are backpropagated through the LMMSE layer into CENet and the earlier transceiver modules. This joint optimization simultaneously improves CE and data detection, while allowing the entire transceiver to adapt in a coherent manner, ensuring that all components operate harmoniously to maximize end-to-end reliability and efficiency.

\section{Simulation Results}
In this section, we present a comprehensive evaluation of the performance of the proposed DL-based end-to-end transceiver. The simulation setup and parameter configurations are detailed in Section \ref{sec.Simulation Details}. Subsequently, the numerical results obtained from simulations are illustrated and analyzed in Section \ref{sec.Numerical Results}, demonstrating the effectiveness and advantages of the proposed scheme under various conditions.
\subsection{Training Details}
\label{sec.Simulation Details}

The entire communication system architecture is implemented using the PyTorch framework. In addition, complex-valued linear and convolutional layers are employed in the system, allowing the network to operate directly in the complex domain, which better reflects the nature of communication signals and enhances both modeling efficiency and representational capacity by preserving amplitude and phase information.

Each module at both the transmitter and receiver is designed with a high degree of flexibility. Except for the CE and detection modules, which must be optimized via NN training, other modules such as constellation mapping/demapping, pilot insertion, and IZak/Zak transforms can be optionally included in the learning process. Specifically, when a module's parameters are set to be trainable, it is optimized through data-driven learning; when fixed, it behaves like a conventional communication block. This allows the system to operate either as a fully end-to-end trainable architecture or as a hybrid model that integrates traditional signal processing components. Detailed designs of each module and the training procedure are provided in Sec.~\ref{sec.Proposed OTFS Transceiver based on DL} and \ref{sec.JCEDD}.

For simulation data generation, in each training epoch, a batch of random binary bit sequences $\mathbf{b} \in \{0,1\}$ with a predefined length is generated. These bits are transmitted through a doubly dispersive channel model that incorporates fractional Doppler effects. After channel, AWGN is added based on the target SNR. At the receiver, the NN attempts to estimate the transmitted bits, with the overall objective being to minimize the BER between the estimated and transmitted sequences.

During training, we employ the Adam optimizer with weight decay to mitigate overfitting, where the weight decay coefficient is tuned between 
$10^{-3}$ and $10^{-4}$ depending on the model configuration. The learning rate is scheduled to gradually decay from $1 \times 10^{-3}$ to $1 \times 10^{-4}$. Training is conducted for 1500 epochs, with a batch size of 16. Model performance is validated every 10 epochs. 
For the neural architectures, when the OTFS grid size is set to $[M,N]=[8,8]$, the numbers of downsampling and upsampling blocks are configured as $L=2$ and $K=5$, respectively, ensuring that the decoder output matches the dimension of the DD-domain equivalent channel matrix $\hat{\mathbf{G}}$. For $[M,N]=[16,16]$, the values of $L$ and $K$ are adjusted accordingly to maintain dimensional consistency.

A detailed list of other system-related hyperparameters is provided in Table \ref{tab.parameter}. Note that unless otherwise specified, the default simulation parameters are: pilot power $E_p$ is set to 10, and the OTFS grid size $[M,N]$ is [8,8]. The main complexity lies in the offline training, while online inference is of manageable complexity.

\begin{table}[t]
\centering
\renewcommand{\arraystretch}{1.2}
\setlength{\tabcolsep}{5pt}
\caption{Simulation parameters.}
\label{tab.parameter}
\begin{tabular}{l l | l l}
\toprule
Parameter & Value  & Parameter & Value  \\
\midrule
Pilot power ($E_p$)         & $5,10$       & Data average power ($E_d$)  & $1$ \\
SNR (dB)                    & $[0,20]$     & Length of CP ($L_{\texttt{CP}}$)     & $4$ \\
Delay bins ($M$)            & $8,16$       & No. of Subcarriers ($N$)            & $8,16$ \\
Modulation order ($R$)      & $2,4$        & Subcarrier spacing ($\Delta f$)    & $15\text{kHz}$ \\
\bottomrule
\end{tabular}
\end{table}

\subsection{Numerical Results}
\label{sec.Numerical Results}
In this section, we conduct simulation experiments to evaluate the performance of each module under end-to-end joint optimization. We omit the performance of  constellation mapping module, as  noticeable constellation point rotations and  shifts compared to traditional designs during joint training have been widely reported in the literature \cite{hu2024end,luo2025bedrock}. 

\begin{figure}[t]
    \centering
    \includegraphics[width=0.9\linewidth]{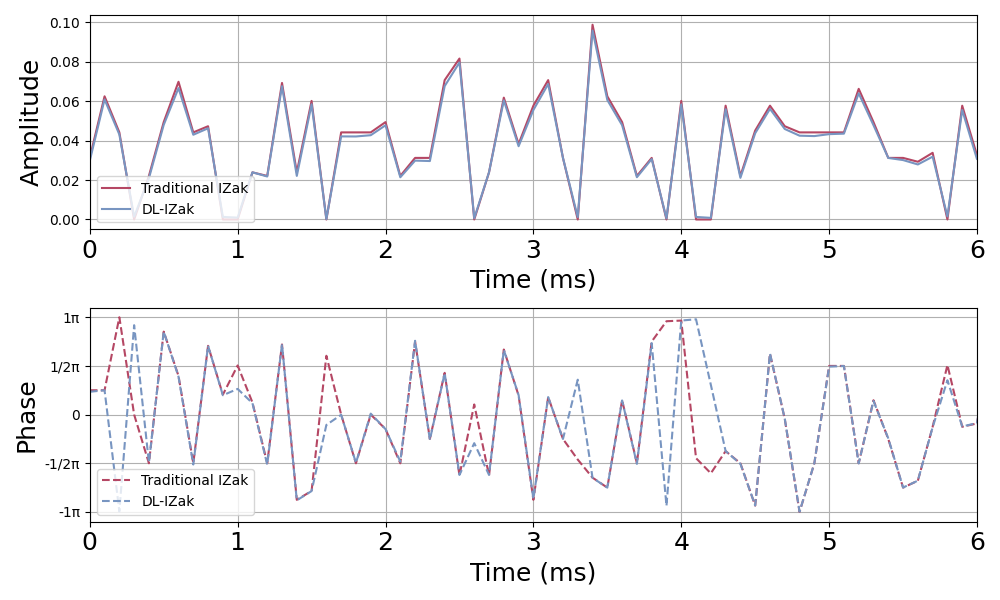}
    \caption{Comparison between the traditional IZak and the proposed DL-IZak method.}
    \label{fig:ISFFT}
\end{figure}

For the IZak module, we compare the traditional IZak transform with our proposed design. As shown in Fig. \ref{fig:ISFFT}, the amplitude remains largely consistent between the two approaches, demonstrating the correctness and compatibility of our design with the traditional method. Meanwhile, noticeable changes in phase indicate the design of DL-IZak will  influence the end-to-end  performance, as will be demonstrated later via simulation results.

\begin{figure}[t]
    \centering
    \includegraphics[width=\linewidth]{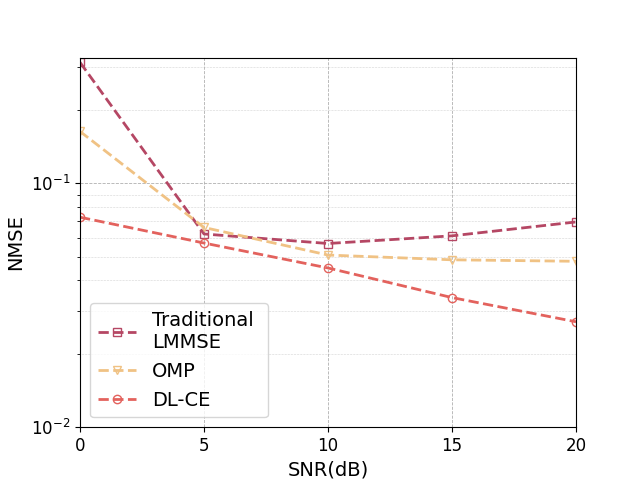}
    \caption{NMSE performance of the proposed DL-CE method, compared with traditional OMP and LMMSE estimators.}
    \label{fig:NMSE}
\end{figure}

For CE, we first evaluate the effectiveness of the proposed pilot separation strategy followed by CENet. This is done by measuring the normalized mean squared error (NMSE) between the estimated equivalent channel and the ground truth, defined as
\begin{align}
    \text{NMSE} = \frac{\mathbb{E}[||\mathbf{G}-\hat{\mathbf{G}}||^2_2]}{\mathbb{E}[||\mathbf{G}||^2_2]}.
\end{align}
We compare the proposed channel estimation method with two conventional baselines: the OMP algorithm and the LMMSE estimator. Note that the LMMSE baseline is evaluated under the embedded-pilot (EP) setting, since in the SP case the interference between pilots and data severely degrades the performance of traditional LMMSE, making it unreliable.
As shown in Fig. \ref{fig:NMSE},
even though EP provides LMMSE with a more favorable operating condition, our method still achieves substantially lower NMSE than both OMP and LMMSE, while using less pilot overhead than the EP-based LMMSE.
The NMSE of the EP-based LMMSE estimator increases at high SNR, likely because it cannot effectively suppress the interference caused by fractional Doppler, which becomes more pronounced as noise diminishes. In contrast, the proposed DL-CE framework consistently outperforms both baselines across the entire SNR range.

To evaluate the effectiveness of our proposed end-to-end joint optimization framework, we compare it against several baseline methods as follows:

\subsubsection{Perfect CSI} A conventional OTFS system employing standard modulation and demodulation techniques, wherein the receiver is assumed to have perfect and instantaneous knowledge of the CSI. Such an assumption enables ideal channel equalization, serving as an upper bound benchmark for system performance in the absence of CE errors.

\subsubsection{DLBench} A DL-based benchmark scheme where transmitter and receiver modules are replaced by a DNN models and trained in an AE manner. 

\subsubsection{Traditional LMMSE} A traditional OTFS system operating under the EP configuration. The EP setting is adopted because, under the SP scheme, the mutual interference between pilots and data severely degrades the performance of conventional LMMSE, making it unable to provide reliable channel estimates.
In the EP setup, the receiver first extracts the pilot symbols embedded in the transmitted frame and performs CE using the LMMSE method. Subsequently, symbol detection is carried out using an LMMSE detector based on the estimated channel.

\subsubsection{Traditional OMP} A conventional OTFS system configured with the SP scheme, where the receiver employs the orthogonal matching pursuit (OMP) algorithm for CE. After CE, LMMSE symbol detection is conducted based on the estimated channel state.

\begin{figure}[t]
    \centering
    \includegraphics[width=\linewidth]{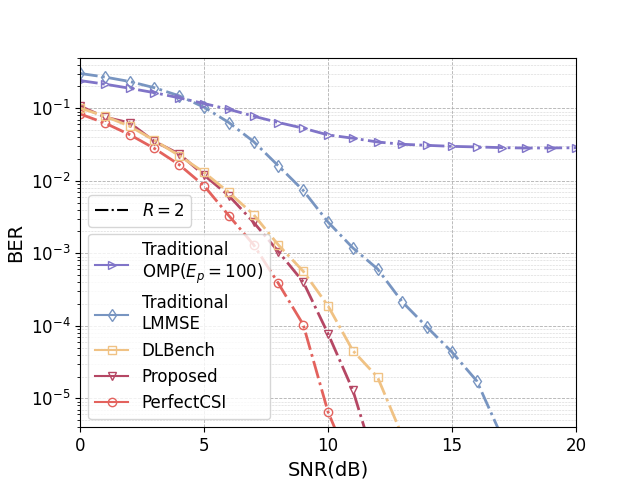}
    \caption{Comparison of BER between the proposed end-to-end method and the comparative method with modulation order $R = 2$.}
    \label{fig:SNR_bpsk}
\end{figure}

\begin{figure}[t]
    \centering
    \includegraphics[width=\linewidth]{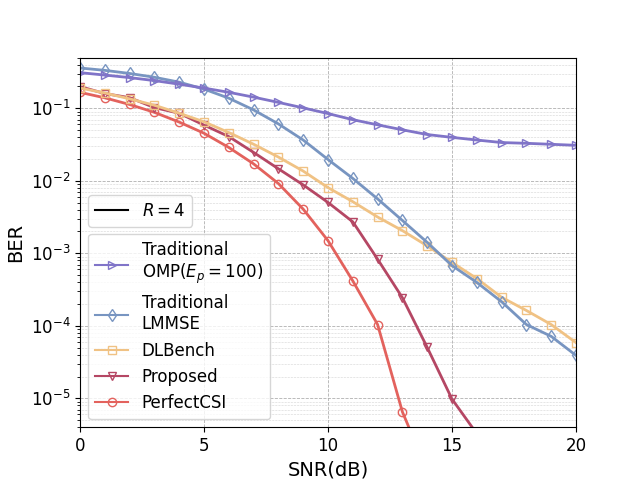}
    \caption{Comparison of BER between the proposed end-to-end method and the comparative method with modulation order $R = 4$.}
    \label{fig:SNR_qpsk}
\end{figure}

The comparison of BER performance under different SNR values for the aforementioned methods is presented in Fig.~\ref{fig:SNR_bpsk} and Fig.~\ref{fig:SNR_qpsk}, where Fig.~\ref{fig:SNR_bpsk} corresponds to a modulation order of 2, i.e., BPSK, and Fig.~\ref{fig:SNR_qpsk} corresponds to a modulation order of 4, i.e., QPSK.

As shown in the figures, our proposed method achieves BER performance close to the Perfect CSI  and outperforms DLBench, highlighting the benefit of joint end-to-end optimization. Furthermore, unlike DLBench, which relies on a single large network to perform CE and detection, our method explicitly estimates the channel, offering improved flexibility. For example, if accurate CSI is available through alternative means, our framework can directly utilize it to enhance detection performance.
In the OMP method, we observe that the BER drops initially but saturates around $3\times 10^{-2}$ even when the pilot power is set to 100. This is likely due to the method’s inability to effectively suppress interference between pilot and data symbols, leading to inaccurate support detection in the DD domain.
Regarding traditional LMMSE, the EP-based approach typically achieves better performance than the SP-based method, as it avoids interference between pilot and data symbols. However, its main limitation lies in the overhead introduced by the guard interval. Specifically, the SP configuration incurs an overhead of 
\begin{align} 
    \eta_\textbf{SP} = \frac{1}{MN},
\end{align}
while the EP configuration results in a higher overhead of 
\begin{align}
    \eta_\textbf{EP} = \frac{(2l_{max}+1)(4k_{max}+1)}{MN}.
\end{align}
A detailed comparison of the overhead between SP and EP configurations is summarized in Table \ref{tab.overhead}.
In our simulations, the proposed end-to-end method outperforms the traditional LMMSE, further demonstrating the effectiveness of our design.

\begin{table}[t]
\renewcommand{\arraystretch}{1.45}
\caption{Overhead comparison between SP and EP.}
\centering
\begin{tabular}{c|c|c|c|c}
\hline
\makecell{} & \textbf{$\mathbf{8 \times8}$} & \textbf{$\mathbf{16 \times 16}$} & \textbf{$\mathbf{32 \times 32}$} & \textbf{$\mathbf{64 \times64}$} \\
\hline
\textbf{EP} & $87.5\%$ & $43.75\%$ & $11.62\%$ & $2.91\%$ \\
\hline
\textbf{SP} & $1.56\%$ & $0.39\%$ & $0.098\%$ & $0.024\%$ \\
\hline
\end{tabular}

\label{tab.overhead}
\end{table}

\begin{figure}[t]
    \centering
    \includegraphics[width=\linewidth]{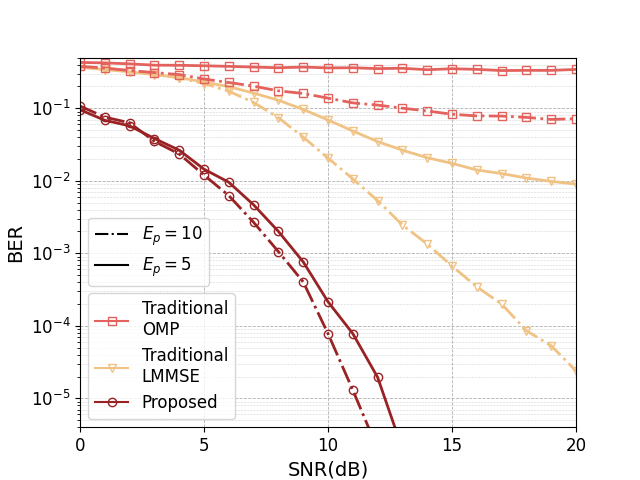}
    \caption{Comparison of BER for different methods at a pilot power $E_p =5, 10$ with modulation order $R=2$.}
    \label{fig:power_bpsk}
\end{figure}

\begin{figure}[t]
    \centering
    \includegraphics[width=\linewidth]{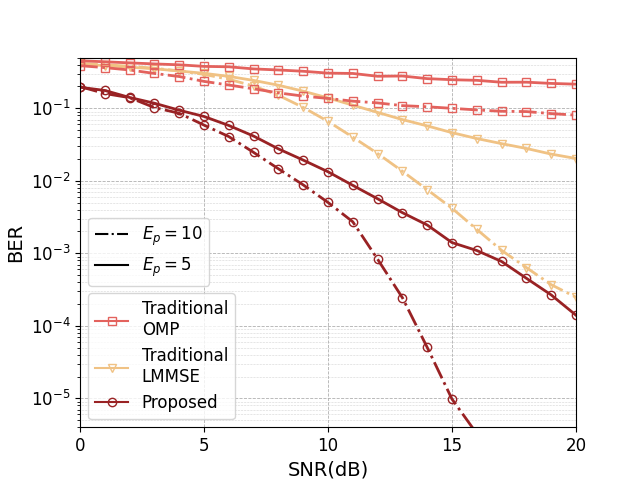}
    \caption{Comparison of BER for different methods at a pilot power $E_p =5, 10$ with modulation order $R=4$.}
    \label{fig:power_qpsk}
\end{figure}

Subsequently, we evaluate the BER performance of our proposed method alongside the benchmark algorithms traditional OMP and traditional LMMSE under varying pilot power levels. The corresponding results are illustrated in Fig. \ref{fig:power_bpsk} and Fig. \ref{fig:power_qpsk}, where Fig. \ref{fig:power_bpsk} represents the scenario with a modulation order of 2, i.e., BPSK, and Fig. \ref{fig:power_qpsk} corresponds to a modulation order of 4, i.e., QPSK. The simulation results clearly demonstrate that increasing the pilot power significantly enhances the receiver’s ability to distinguish pilot symbols. This leads to more accurate CE, which, in turn, improves the overall data detection performance and reduces BER.
Nevertheless, while allocating higher power to pilot symbols can improve CE accuracy, it may also increase the PAPR of the transmitted signal. In contrast, our proposed method achieves better performance than traditional approaches even when using lower pilot power, demonstrating its ability to reduce PAPR while maintaining accurate channel estimation.

\begin{figure}[t]
    \centering
    \includegraphics[width=\linewidth]{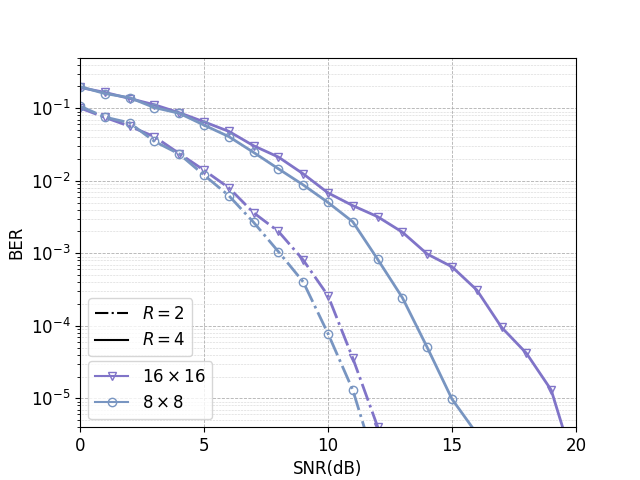}
    \caption{The comparison of BER under different numbers of delay bins and subcarriers, i.e., [8,8] and [16,16], for the proposed method.}
    \vspace{-0.3 cm}
    \label{fig:grid}
\end{figure}

Finally, we extend our evaluation to a larger configuration with the number of delay bins and subcarriers set to 
[16,16]. To effectively handle this increased dimensionality, the architecture of the CENet is suitably adjusted to accommodate the expanded input and output sizes. Subsequently, the entire end-to-end system is retrained from scratch to adapt to the new configuration.
The training results for the proposed method under different delay bin and subcarrier configurations are presented in Fig. \ref{fig:grid}. As depicted, our proposed design demonstrates strong generalization capabilities across varying grid dimensions. Although the performance under the $[M,N]=[16,16]$ setting is slightly inferior to that of the smaller $[8,8]$ configuration, the system still attains satisfactory detection accuracy and overall effectiveness. This observed performance gap is mainly attributed to the substantially larger equivalent channel matrix in the [16,16] case, which increases the complexity of the learning task. As a result, the NN requires enhanced modeling capacity and training effort to fully capture the channel characteristics at this higher dimension.

\section{Conclusions}
In this work, we presented a flexible and modular DL-based OTFS system architecture that enables end-to-end training while maintaining compatibility with traditional communication modules. By allowing each component, such as constellation mapping, pilot insertion and IZak transforms, to be either fixed or jointly optimized, our framework supports a wide range of design choices, from partially learned to fully end-to-end systems.
We evaluated the proposed system under SP configuration in the presence of challenging fractional Doppler channels. Experimental results demonstrate that our method consistently outperforms conventional baselines, achieving BER performance close to that with perfect channel knowledge.
Our approach generalizes well across varying OTFS grid sizes, maintaining stable performance even when transitioning from [8,8] to [16,16] configurations, albeit with a slight performance drop due to the increased complexity of the equivalent channel matrix. The adoption of complex-valued NN layers enhances the model’s ability to handle amplitude and phase information natively, improving representational efficiency in the complex domain.


\bibliographystyle{ieeetr}	
\bibliography{ref}

@ARTICLE{9508932,
  author={Wei, Zhiqiang and Yuan, Weijie and Li, Shuangyang and Yuan, Jinhong and Bharatula, Ganesh and Hadani, Ronny and Hanzo, Lajos},
  journal={IEEE Wireless Commun.}, 
  title={Orthogonal Time-Frequency Space Modulation: A Promising Next-Generation Waveform}, 
  year={2021},
  volume={28},
  number={4},
  pages={136-144},
  doi={10.1109/MWC.001.2000408}}

@ARTICLE{9785832,
  author={Wang, Xueyang and Shen, Wenqian and Xing, Chengwen and An, Jianping and Hanzo, Lajos},
  journal={IEEE Trans. Commun.}, 
  title={Joint Bayesian Channel Estimation and Data Detection for {{OTFS}} Systems in {LEO} Satellite Communications}, 
  year={2022},
  volume={70},
  number={7},
  pages={4386-4399},
  doi={10.1109/TCOMM.2022.3179389}}

@ARTICLE{10370744,
  author={Liang, Yu and Fan, Pingzhi and Wang, Qianli and He, Xiaolin},
  journal={IEEE Trans. Wireless Commun.}, 
  title={Two-Dimensional Delay-Doppler Pilots and Channel Estimation for Multi-Antenna {{OTFS}} in Doubly Dispersive Channels}, 
  year={2024},
  volume={23},
  number={7},
  pages={7612-7623},
  doi={10.1109/TWC.2023.3342877}}

@ARTICLE{10637960,
  author={He, Xiaolin and Yuan, Weijie and Fan, Pingzhi},
  journal={IEEE Trans. Wireless Commun.}, 
  title={On the Pilot-Aided Channel Estimation for Windowed {{OTFS}} With Data Interference in Rapidly Time-Varying Channels}, 
  year={2024},
  volume={23},
  number={11},
  pages={16359-16374},
  doi={10.1109/TWC.2024.3440586}}

@ARTICLE{10038843,
  author={Liu, Wei and Zou, Liyi and Bai, Baoming and Sun, Teng},
  journal={China Commun.}, 
  title={Low {PAPR} channel estimation for {{OTFS}} with scattered superimposed pilots}, 
  year={2023},
  volume={20},
  number={1},
  pages={79-87},
  doi={10.23919/JCC.2023.01.007}}

@ARTICLE{9285313,
  author={Qu, Huiyang and Liu, Guanghui and Zhang, Lei and Wen, Shan and Imran, Muhammad Ali},
  journal={IEEE Trans. Commun.}, 
  title={Low-Complexity Symbol Detection and Interference Cancellation for {{OTFS}} System}, 
  year={2021},
  volume={69},
  number={3},
  pages={1524-1537},
  doi={10.1109/TCOMM.2020.3043007}}

@ARTICLE{9738478,
  author={Wei, Zhiqiang and Yuan, Weijie and Li, Shuangyang and Yuan, Jinhong and Ng, Derrick Wing Kwan},
  journal={IEEE Trans. Wireless Commun.}, 
  title={Off-Grid Channel Estimation With Sparse Bayesian Learning for {{OTFS}} Systems}, 
  year={2022},
  volume={21},
  number={9},
  pages={7407-7426},
  doi={10.1109/TWC.2022.3158616}}

@ARTICLE{10506450,
  author={Tang, Meng and Wang, Hao and Yuan, Zongming and Yuan, Jinhong},
  journal={IEEE Trans. Wireless Commun.}, 
  title={A Novel Off-Grid Channel Estimation With Fast BCS Using {LSM} Prior for {{OTFS}} Modulation}, 
  year={2024},
  volume={23},
  number={9},
  pages={12157-12171},
  doi={10.1109/TWC.2024.3388449}}

@ARTICLE{10685072,
  author={Liu, Yu and Chen, Ming and Pan, Cunhua and Gong, Tantao and Yuan, Jinhong and Wang, Jiangzhou},
  journal={IEEE J. Sel. Areas Commun.}, 
  title={{{OTFS}} Versus {OFDM}: Which is Superior in Multiuser {LEO} Satellite Communications}, 
  year={2025},
  volume={43},
  number={1},
  pages={139-155},
  doi={10.1109/JSAC.2024.3460060}}

@ARTICLE{8424569,
  author={Raviteja, P. and Phan, Khoa T. and Hong, Yi and Viterbo, Emanuele},
  journal={IEEE Trans. Wireless Commun.}, 
  title={Interference Cancellation and Iterative Detection for Orthogonal Time Frequency Space Modulation}, 
  year={2018},
  volume={17},
  number={10},
  pages={6501-6515},
  doi={10.1109/TWC.2018.2860011}}

@ARTICLE{9973328,
  author={Yuan, Zongming and Tang, Meng and Chen, Jianhua and Fang, Taibin and Wang, Hao and Yuan, Jinhong},
  journal={IEEE Trans. Veh. Technol.}, 
  title={Low Complexity Parallel Symbol Detection for {{OTFS}} Modulation}, 
  year={2023},
  volume={72},
  number={4},
  pages={4904-4918},
  doi={10.1109/TVT.2022.3227282}}

@ARTICLE{10105487,
  author={Li, Shuo and Ding, Chao and Xiao, Lixia and Zhang, Xufan and Liu, Guanghua and Jiang, Tao},
  journal={IEEE Trans. Veh. Technol.}, 
  title={Expectation Propagation Aided Model Driven Learning for {{OTFS}} Signal Detection}, 
  year={2023},
  volume={72},
  number={9},
  pages={12407-12412},
  doi={10.1109/TVT.2023.3268231}}

@ARTICLE{10423015,
  author={Yue, Yang and Shi, Jia and Li, Zan and Hu, Junfan and Tie, Zhuangzhuang},
  journal={IEEE Commun. Lett.}, 
  title={Model-Driven Deep Learning Assisted Detector for {{OTFS}} With Channel Estimation Error}, 
  year={2024},
  volume={28},
  number={4},
  pages={842-846},
  doi={10.1109/LCOMM.2024.3362970}}

@ARTICLE{10439989,
  author={Mattu, Sandesh Rao and Chockalingam, A.},
  journal={IEEE Wireless Commun. Lett.}, 
  title={Learning in Time-Frequency Domain for Fractional Delay-Doppler Channel Estimation in {{OTFS}}}, 
  year={2024},
  volume={13},
  number={5},
  pages={1245-1249},
  doi={10.1109/LWC.2024.3367112}}

@ARTICLE{10816508,
  author={Qing, Chaojin and Liu, Zhiying and Ling, Guowei and Hu, Wenquan and Du, Pengfei},
  journal={IEEE Trans. Veh. Technol.}, 
  title={Channel Estimation in {{OTFS}} Systems by Leveraging Differential Modulation}, 
  year={2025},
  volume={74},
  number={5},
  pages={6907-6918},
  doi={10.1109/TVT.2024.3522940}}

@ARTICLE{10138432,
  author={Guo, Lin and Gu, Peng and Zou, Jun and Liu, Guangzu and Shu, Feng},
  journal={IEEE Trans. Veh. Technol.}, 
  title={{DNN}-Based Fractional Doppler Channel Estimation for {{OTFS}} Modulation}, 
  year={2023},
  volume={72},
  number={11},
  pages={15062-15067},
  doi={10.1109/TVT.2023.3280901}}

@ARTICLE{10747184,
  author={Ying, Daidong and Ye, Feng},
  journal={IEEE Trans. Veh. Technol.}, 
  title={Deep Learning Supported Path Prediction and Channel Estimation for {MIMO}-{{OTFS}} System With High Delay Resolution}, 
  year={2025},
  volume={74},
  number={3},
  pages={3584-3597},
  doi={10.1109/TVT.2024.3493921}}

@ARTICLE{10856399,
  author={Qi, Shuyuan and Wang, Qianli and Ma, Zheng},
  journal={IEEE Trans. Veh. Technol.}, 
  title={Deep Residual Attention Network for {{OTFS}} Channel Estimation}, 
  year={2025},
  volume={74},
  number={6},
  pages={9834-9839},
  doi={10.1109/TVT.2025.3534796}}

@ARTICLE{10623419,
  author={Xu, Jiarui and Said, Karim and Zheng, Lizhong and Liu, Lingjia},
  journal={IEEE Trans. Wireless Commun.}, 
  title={{2D-RC}: Two-Dimensional Neural Network Approach for {{OTFS}} Symbol Detection}, 
  year={2024},
  volume={23},
  number={12},
  pages={17825-17840},
  doi={10.1109/TWC.2024.3407715}}

@ARTICLE{10552800,
  author={Singh, Amit and Sharma, Sanjeev and Sharma, Mohit and Deka, Kuntal and da Costa, Daniel B.},
  journal={IEEE Wireless Commun. Lett.}, 
  title={Autoencoder-Based End-to-End {{OTFS}} System Design With Hardware Impairments}, 
  year={2024},
  volume={13},
  number={8},
  pages={2285-2289},
  doi={10.1109/LWC.2024.3412240}}

@ARTICLE{10226266,
  author={Tek, Yusuf Islam and Dogukan, Ali Tugberk and Basar, Ertugrul},
  journal={IEEE Commun. Lett.}, 
  title={Autoencoder-Based Enhanced Orthogonal Time Frequency Space Modulation}, 
  year={2023},
  volume={27},
  number={10},
  pages={2628-2632},
  doi={10.1109/LCOMM.2023.3307423}}

@ARTICLE{9448141,
  author={Zhao, Zhongyuan and Vuran, Mehmet Can and Guo, Fujuan and Scott, Stephen D.},
  journal={IEEE J. Sel. Areas Commun.}, 
  title={Deep-Waveform: A Learned OFDM Receiver Based on Deep Complex-Valued Convolutional Networks}, 
  year={2021},
  volume={39},
  number={8},
  pages={2407-2420},
  doi={10.1109/JSAC.2021.3087241}}

@article{luo2025bedrock,
  title={Bedrock Models in Communication and Sensing: Advancing Generalization, Transferability, and Performance},
  author={Luo, Cheng and Xiang, Luping and Hu, Jie and Yang, Kun},
  journal={arXiv preprint arXiv:2503.08220},
  year={2025}
}

@ARTICLE{10769778,
  author={Rou, Hyeon Seok and de Abreu, Giuseppe Thadeu Freitas and Choi, Junil and González G., David and Kountouris, Marios and Guan, Yong Liang and Gonsa, Osvaldo},
  journal={IEEE Signal Processing Magazine}, 
  title={From Orthogonal Time–Frequency Space to Affine Frequency-Division Multiplexing: A comparative study of next-generation waveforms for integrated sensing and communications in doubly dispersive channels}, 
  year={2024},
  volume={41},
  number={5},
  pages={71-86},
  doi={10.1109/MSP.2024.3422653}}

@ARTICLE{11071963,
  author={Gao, Yuan and Xu, Xinchen and Jin, Yanliang and Yuan, Weijie and Zhang, Jie and Xu, Shugong},
  journal={IEEE Internet of Things Journal}, 
  title={Joint Channel Estimation and Data Detection for {OTFS} Systems: A Lightweight Deep Learning Framework With a Novel Data Augmentation Method}, 
  year={2025},
  volume={12},
  number={18},
  pages={38464-38481},
  doi={10.1109/JIOT.2025.3586012}}

@ARTICLE{11207204,
  author={Cheng, Jiaming and Chen, Wei and Yang, Bowen and Ai, Bo},
  journal={Journal of Communications and Information Networks}, 
  title={{LLM}-Enhanced Pilot-Free {OTFS} Transceiver for High-Mobility Communications}, 
  year={2025},
  volume={10},
  number={3},
  pages={215-223},
  doi={10.23919/JCIN.2025.11207204}}

@INPROCEEDINGS{9473532,
  author={Hashimoto, Noriyuki and Osawa, Noboru and Yamazaki, Kosuke and Ibi, Shinsuke},
  booktitle={2021 IEEE International Conference on Communications Workshops (ICC Workshops)}, 
  title={Channel Estimation and Equalization for CP-OFDM-based {OTFS} in Fractional Doppler Channels}, 
  year={2021},
  volume={},
  number={},
  pages={1-7},
  doi={10.1109/ICCWorkshops50388.2021.9473532}}

@misc{ronneberger2015unetconvolutionalnetworksbiomedical,
      title={{U-Net}: Convolutional Networks for Biomedical Image Segmentation}, 
      author={Olaf Ronneberger and Philipp Fischer and Thomas Brox},
      year={2015},
      eprint={1505.04597},
      archivePrefix={arXiv},
      primaryClass={cs.CV},
      url={https://arxiv.org/abs/1505.04597}
}

@ARTICLE{11263980,
  author={Zhang, Xiaoqi and Ni, Zhitong and Yuan, Weijie and Andrew Zhang, J. and Quek, Tony Q. S.},
  journal={IEEE Transactions on Communications}, 
  title={Deep Learning-based {OTFS} Channel Estimation and Symbol Detection with Plug-and-Play Framework}, 
  year={2025},
  volume={},
  number={},
  pages={1-1},
  doi={10.1109/TCOMM.2025.3636093}}

@ARTICLE{9508141,
  author={Xiang, Luping and Liu, Yusha and Yang, Lie-Liang and Hanzo, Lajos},
  journal={IEEE Transactions on Vehicular Technology}, 
  title={Gaussian Approximate Message Passing Detection of Orthogonal Time Frequency Space Modulation}, 
  year={2021},
  volume={70},
  number={10},
  pages={10999-11004},
  doi={10.1109/TVT.2021.3102673}}

@ARTICLE{10453468,
  author={Xu, Chao and Xiang, Luping and Sugiura, Shinya and Maunder, Robert G. and Yang, Lie-Liang and Niyato, Dusit and Li, Geoffrey Ye and Schober, Robert and Hanzo, Lajos},
  journal={IEEE Transactions on Wireless Communications}, 
  title={Noncoherent Orthogonal Time Frequency Space Modulation}, 
  year={2024},
  volume={23},
  number={8},
  pages={10072-10090},
  doi={10.1109/TWC.2024.3368406}}

@ARTICLE{10454001,
  author={Li, Shuo and Xiao, Lixia and He, Chunlin and Li, Liuke and Jiang, Tao},
  journal={IEEE Transactions on Vehicular Technology}, 
  title={Approximated Expectation Propagation Assisted Decentralized Signal Detection for Uplink Massive {MIMO}-{OTFS} Systems}, 
  year={2024},
  volume={73},
  number={7},
  pages={10280-10286},
  doi={10.1109/TVT.2024.3371453}}

@ARTICLE{10404096,
  author={Xiang, Luping and Xu, Ke and Hu, Jie and Masouros, Christos and Yang, Kun},
  journal={IEEE Internet of Things Journal}, 
  title={Robust NOMA-Assisted {OTFS}-ISAC Network Design With 3-D Motion Prediction Topology}, 
  year={2024},
  volume={11},
  number={9},
  pages={15909-15918},
  doi={10.1109/JIOT.2024.3352391}}

@article{liu2022learning,
  title={Learning how to transfer from uplink to downlink via hyper-recurrent neural network for FDD massive {MIMO}},
  author={Liu, Yusha and Simeone, Osvaldo},
  journal={IEEE Transactions on Wireless Communications},
  volume={21},
  number={10},
  pages={7975--7989},
  year={2022},
  publisher={IEEE}
}

@article{hu2024end,
  title={End-to-End Design of Polar Coded Integrated Data and Energy Networking},
  author={Hu, Jie and Cui, Jingwen and Xiang, Luping and Yang, Kun},
  journal={IEEE Transactions on Communications},
  volume={72},
  number={11},
  pages={7017--7031},
  year={2024},
  publisher={IEEE}
}

\end{document}